\documentclass[twocolumn, longbib]{aastex7}

\usepackage{microtype}
\usepackage{url}
\usepackage{amsmath}
\usepackage{mathtools}
\usepackage{esint}
\usepackage{amssymb}
\usepackage{multirow}
\usepackage{graphicx}
\usepackage{scalerel}
\usepackage{calc}
\usepackage{etoolbox}
\usepackage{marginnote}
\usepackage{nicefrac}
\usepackage{tabstackengine}
\usepackage{diagbox}
\usepackage[makeroom]{cancel}
\usepackage{mathdots}
\usepackage{bbm}
\usepackage{booktabs}
\usepackage{xspace}
\usepackage{upgreek}
\usepackage[T1]{fontenc} \usepackage{textcomp}
\usepackage{ifxetex}
\ifxetex
\usepackage{fontspec}
\defaultfontfeatures{Extension = .otf}
\fi
\usepackage{fontawesome}
\usepackage{listings}
\usepackage{mathtools}
\stackMath

\newcommand{\pycodelink}[1]{\href{#1}{\codeicon}\@\xspace}
\newcommand{\jlcodelink}[1]{\href{#1}{\codeicon}\@\xspace}
\newcommand{\animlink}[1]{\href{#1}{\animicon}\@\xspace}
\newcommand{\prooflink}[1]{\href{#1}{\raisebox{-0.1em}{\prooficon}}}

\renewcommand{\eqref}[1]{\ref{eq:#1}}

\definecolor{linkcolor}{rgb}{0.1216,0.4667,0.7059}
\newcommand{\codeicon}{{\color{linkcolor}\faFileCodeO}}
\newcommand{\prooficon}{{\color{linkcolor}\faPencilSquareO}}
\newcommand{\animicon}{{\color{linkcolor}\faPlayCircle}}

\newtagform{eqtag}[]{(}{)}
\newcommand{\currentlabel}{None}

\newenvironment{proof*}[1]{\ifstrempty{#1}{\renewtagform{eqtag}[]{\raisebox{-0.1em}{{\color{red}\faPencilSquareO}}\,(}{)}}{\renewtagform{eqtag}[]{\prooflink{#1}\,(}{)}}\usetagform{eqtag}\renewcommand{\currentlabel}{#1}
\equation }{\endequation \renewtagform{eqtag}[]{(}{)}\usetagform{eqtag}\message{<<<\currentlabel: \theequation>>>}
}

\newcommand{\T}{\ensuremath{\mathrm{T}}}

\newcommand{\x}{\ensuremath{\mbox{$x$}}}
\newcommand{\y}{\ensuremath{\mbox{$y$}}}
\newcommand{\z}{\ensuremath{\mbox{$z$}}}

\DeclareMathAlphabet\mathbfcal{OMS}{cmsy}{b}{n}

\definecolor{dim}{rgb}{0.8,0.8,0.8}
\newcolumntype{L}[1]{>{\raggedright\let\newline\\\arraybackslash\hspace{0pt}}m{#1}}
\usepackage{listings}
\definecolor{codegreen}{rgb}{0,0.6,0}
\definecolor{codegray}{rgb}{0.5,0.5,0.5}
\definecolor{codepurple}{rgb}{0.58,0,0.82}
\definecolor{backcolour}{rgb}{0.95,0.95,0.95}
\lstdefinestyle{mystyle}{
    backgroundcolor=\color{backcolour},
    commentstyle=\color{codegreen},
    keywordstyle=\color{magenta},
    numberstyle=\tiny\color{codegray},
    stringstyle=\color{codepurple},
    basicstyle=\small\ttfamily,
    breakatwhitespace=false,
    breaklines=true,
    captionpos=b,
    keepspaces=true,
    numbers=left,
    numbersep=5pt,
    showspaces=false,
    showstringspaces=false,
    showtabs=false,
    tabsize=2,
    aboveskip=1em,
    belowskip=1em,
    keywords=[2]{map},
    keywordstyle=[2]{\color{black!80!black}},
}
\newcommand{\jorbit}{\texttt{jorbit}\@\xspace}

\newcommand{\assist}{\texttt{assist}\@\xspace}
\newcommand{\rebound}{\texttt{REBOUND}\@\xspace}

\newcommand{\jax}{\texttt{JAX}\@\xspace}

\submitjournal{PSJ}
\shorttitle{jorbit}
\graphicspath{{./}{}}

\usepackage{amssymb}
\usepackage{lipsum}
\usepackage{bm}
\usepackage{listings}
\usepackage{xcolor}
\usepackage{amsmath}

\begin{document}

\title{A High-Precision, Differentiable Code for Solar System Ephemerides}

\correspondingauthor{Ben Cassese}
\email{b.c.cassese@columbia.edu}

\author[0000-0002-9544-0118]{Ben Cassese}
\affiliation{Dept. of Astronomy, Columbia University, 550 W 120th Street, New York NY 10027, USA}
\affiliation{Department of Astronomy, Yale University, New Haven, CT 06511, USA}
\email{b.c.cassese@columbia.edu}

\author[0000-0002-7670-670X]{Malena Rice}
\affiliation{Department of Astronomy, Yale University, New Haven, CT 06511, USA}
\email{malena.rice@yale.edu}

\author[0000-0003-0834-8645]{Tiger Lu}
\affiliation{Department of Astronomy, Yale University, New Haven, CT 06511, USA}
\email{tiger.lu@yale.edu}

\begin{abstract}

We present \jorbit, a python/\jax library designed to enable modern data-driven numerical studies of the solar system. Written entirely in \jax, an auto-differentiable and optionally GPU accelerated language behind many current large-scale machine learning efforts, \jorbit includes an independent implementation of \rebound's IAS15 integrator and the ability to parse precomputed ephemerides such as the JPL DE series. In its default behavior, \jorbit maintains $\sim$1\,mas agreement with JPL Horizons on $\sim$decade timescales for typical main-belt asteroids, enabling it to fully capitalize on high-precision astrometry and ranging data. We include details of the code's implementation and several worked examples, including illustrations of \jorbit's ability to simulate N-body systems, forward model astrometric data, fit orbits, replicate the Minor Planet Center's ``MPChecker'' service, and contribute to modeling the effect of minor planets on stellar light curves.

\end{abstract}


\section{Introduction} \label{sec:intro}

Determining the orbits of objects within the solar system is one of the oldest tasks of astronomy, itself one of the oldest natural sciences. Ancient cultures could predict the motions of the planets far into the future; the modern form of the practice, assigning a set of six Keplerian orbital elements tagged to a particular epoch, has itself been a common task for more than 200 years \citep{gauss_1809}.

In light of these centuries of work, it is perhaps surprising that considerable efforts have been expended to further develop orbit fitting routines in recent decades. Beginning in the 1960s, the geometric methods first invented by Gauss and refined over the years were translated into computer software \citep{herget_1965, marsden_1985}. In the 1990s, a generation of new routines was developed and tentative steps were taken to move beyond single-point least-squares estimates \citep{muinonen_1993, milani_1999}. The 2000s-2010s saw many of these routines codified into open source software packages, and by the mid 2020s, would-be orbit fitters had a multitude of methods and implementations to choose from.

In \citet{bernstein_2000}, a publication that introduced a new orbit fitting code, the authors quipped: ``Workstations are even fast enough to bound the uncertainty region with a brute-force sampling of the six-dimensional orbit space in many cases. Why then should we bother developing another orbit-fitting technique?''. Two decades and several additional software packages later, \citet{holman_assist_2023} similarly ask in their own code release paper: ``This [plethora of orbit fitting routines] raises an obvious question: Why develop another such package?'' Considering that this manuscript adds to the relative bloat and introduces yet another orbit fitting software, it is only fair that we consider this question ourselves.

\jorbit was designed with three primary objectives in mind, each of which were unmet by the current suite of available orbit fitting routines at the initiation of our software construction. First, we wanted to entirely abandon any Keplerian or linearized assumptions and to base all fits on N-body simulations that include as many known perturbative effects as possible. Second, we wanted the ability to fit these dynamical models using modern (potentially gradient-based) MCMC methods that so far have not been popular in the minor planet literature, but that can enable inference of very high-dimensional, complex problems, such as the gravitational interaction of multiple bodies. Though in principle it is possible to compute these gradients manually, a practice known in the dynamics literature as creating ``variational equations'' \citep[e.g.][]{rein_2nd_order_2016, holman_assist_2023}, we aimed for a more flexible approach based on autodifferentiation that allows for more rapid model construction/iteration as well as easier access to higher-order derivatives. Finally, we aimed to ensure that each step of using the software was user-friendly, from installation to exploratory analysis to the creation of final science-quality results. To this end, we strove to practice proper open-source principles and to maintain active online documentation and issue trackers.

We have achieved each of these objectives, but happily, were not the only project to attempt the first and last. Early into the preparation of this software, \citet{holman_assist_2023} released \assist, a new orbit integration package developed within the \rebound \citep{rein_rebound_2012} ecosystem. This powerful tool nicely complements \jorbit's capabilities: both packages are based off of the same IAS15 integration algorithm \citep{everhart_1985, rein_ias15_2015}, though they are written in different languages (\texttt{C99} vs. python/\jax{}), share no exact source code (though as discussed in Sec. \ref{sub:integrator}, one of our integrators is modeled as closely as possible to the \rebound source and shares the same numerical constants). As of writing, however, \assist does not include any orbit fitting routines or an interface for converting 3D positions to on-sky astrometry, and instead focuses only on high-precision orbit integration. Even more recently, \citet{grss} introduced \texttt{GRSS}, a \texttt{C++}-based implementation of IAS15 that includes numerous orbit fitting routines. To our knowledge, though, \jorbit remains unique in its ability to autodifferentiate any arbitrary user-defined acceleration model, meaning it alone can plug into any gradient-dependent fitting methods such as Hamiltonian Monte Carlo algorithms without significant additional mathematics. This power does come with a penalty, however: \jorbit is consistently slower than \assist for like-for-like simulations, though generally runtimes are within a factor of a few.

In addition to its orbit-fitting capabilities, we built an additional branch of tools into the package: the ability to replicate the Minor Planet Center's (MPC) ``MPChecker'' service\footnote{\href{https://www.minorplanetcenter.net/cgi-bin/checkmp.cgi}{MPChecker Web Service}}. Traditionally, if one wished to ask where a given minor planet might be on a given night, one would turn to a service like the Minor Planet Center or JPL Horizons\footnote{\href{https://ssd.jpl.nasa.gov/horizons/}{JPL Horizons Web Service}}, which can generate a particle-specific ephemeris relatively quickly. However, the \textit{inverse} problem is much more challenging: if one has a patch of sky and wants to know all of the minor planets contained within it at a certain time, they have to use a service like MPChecker or the Virtual Observatory of the IMCEE CNRS\footnote{\href{https://ssp.imcce.fr/webservices/skybot/}{SkyBoT: The Virtual Observatory Sky Body Tracker}} to simulate the positions of \textit{all} known minor planets and check which are contained within the footprint. This is much slower and relies heavily on computations carried out by these remote services, which makes it difficult to create a time series of simulated observations. \jorbit allows users to carry out these computations entirely on their local machines, which unlocks the ability to check time series observations of other science targets (such as transiting exoplanets or transient events) for contamination from minor planets.

Throughout this article, we include links to specific portions of the code base with the following icon: \pycodelink{https://github.com/ben-cassese/jorbit}. Though we expect \jorbit to evolve over time, the links here will be tagged to a frozen version of the code associated with a specific Github commit. Additionally, we have archived version 1.0.0 of the code on Zenodo \citep{zenodo_release}. We dedicate Sec. \ref{sec:code} to a description of the underlying integration models, and Sec. \ref{sec:mpchecker} to a description of the algorithm beneath the MPChecker-related functions; users less interested in the code implementation will be more comfortable in Sec. \ref{sec:examples}, where we present several examples of the software's capabilities. We conclude in Sec. \ref{sec:conclusion} with some thoughts towards future potential package developments.

\section{Model Implementation} \label{sec:code}

\subsection{Philosophy} \label{sub:philosophy}

We aimed for four design requirements when constructing \jorbit:
\begin{itemize}
    \item \jax wherever possible
    \item Flexibility/extensibility in a lower-level interface
    \item Easy-to-use higher level wrappers
    \item Limited scope: do not include fitting routines, only what is necessary to compute a likelihood.
\end{itemize}

Addressing the first point, although \jax is a python package, in practice it is easier to think of it as its own distinct programming language. For a block of code to be eligible for just-in-time (JIT) compilation or GPU dispatch, it generally must be written in the functional programming paradigm and avoid the \jax ``sharp-bits''\footnote{For an exhaustive description of and introduction to JAX, we refer readers to the \href{https://docs.jax.dev/en/latest/notebooks/Common_Gotchas_in_JAX.html}{official documentation}.}. By following \jax best-practices right up until the user interface level, we maximize the amount of code that can leverage its advantages.

Addressing the second point, we designed our integrator to accept any \jax function that computes the instantaneous acceleration felt by a collection of particles. We include several built-in functions that compute gravitational interactions (both Newtonian and parameterized post-Newtonian, see Sec. \ref{sub:accelerations}), the contributions of arbitrary-order gravitational harmonics, and non-gravitational forces \citep{marsden_1973}, but interested users can easily extend beyond these to model more complex dynamics. For example, one could create a time-dependent source of acceleration caused by processes like solar mass loss (e.g., \citet{duncan_1998}). This leaves open the possibility both for individual users to experiment on their own, and for additional acceleration functions to be added to the central repository in the future, much like how \texttt{REBOUNDx} \citep{reboundx} supplements the \rebound ecosystem with more advanced acceleration functions.

For the third, we aimed to reduce the amount of required ``boilerplate'' code as much as possible: for users who just want to integrate a massless particle within the solar system, we created wrappers like the \texttt{Particle} (\pycodelink{https://github.com/ben-cassese/jorbit/blob/907e5b8ccee42479bf4fa67435b5733318d0bd9d/src/jorbit/particle.py}) and \texttt{System} (\pycodelink{https://github.com/ben-cassese/jorbit/blob/907e5b8ccee42479bf4fa67435b5733318d0bd9d/src/jorbit/system.py}) classes that allow for simple simulations/ephemeris generation with just a few lines of python.

Finally, we aimed to keep \jorbit small: every user has their preferred algorithm/framework for fitting a model to data, and our objective is only to provide them with the model and the target function to optimize. We include one built-in function to compute the maximum likelihood estimate of a standard orbit, but leave the rest of the fitting process to the user.

\subsection{Integrator} \label{sub:integrator}

Thanks to the impossibility of an analytic solution to the general N-body problem, any attempt to model the motion of objects in the solar system with high precision must rely on a numerical integrator. The choice of \textit{which} integrator is determined by our specific application: \jorbit operates in a regime where precision is paramount and runtime is a secondary concern. 

To illustrate why, consider \jorbit's conventional use case: \jorbit is designed to simulate the solar system, but more specifically, to fit unknown quantities to \textit{data} of the solar system. If we are willing to consider all quantitative observations of minor planets ever taken as ``data'', our problem consists of simulations that will last at most from the discovery of Ceres in the 19th century to the present. This maximum baseline is only a few hundred dynamical cycles of each target, and this still represents an extreme case compared to the outer solar system. Neptune has only recently completed its first $\sim$165-year orbit since its discovery in 1846, and no objects farther out have been observed over one full orbit. One could imagine wishing to resolve moons or asteroid binaries over similar timescales, but even in these cases we likely need to simulate fewer than $10^6$ total orbits. While it will still be convenient to optimize our runtime performance, our algorithm will not need to withstand the pressure of $>10^9$ dynamical times faced by codes used to model effects like secular interactions and system stability \citep{wisdom_holman_1991, tamayo_spock_2020}.

The positional precision requirements of these short simulations, however, are formidable. The \textit{Gaia} spacecraft routinely observes main-belt asteroids with a single-transit astrometric precision below 1 mas in the along scan direction \citep{gaia_dr2_solar_system_2018, tanga_gaia_dr3_2023}. At a distance of 2 AU, misjudging the position of a target by only $\sim10^2$ meters could lead to an astrometric shift of a comparable size. Since in five years, the length of \textit{Gaia}’s nominal mission, a typical main belt object will travel $\sim10^{12}$\,m, our integration scheme must be accurate to at least within one part in $10^{10}$ to avoid artificially biasing our inference. Radio ranging of planetary science missions provides similarly strict requirements and in some circumstances produces measurements of a probe's line-of-sight distance with sub-meter precision \citep{fienga_2016}.

\begin{figure}[!h]
    \centering
    \includegraphics[width=\columnwidth]{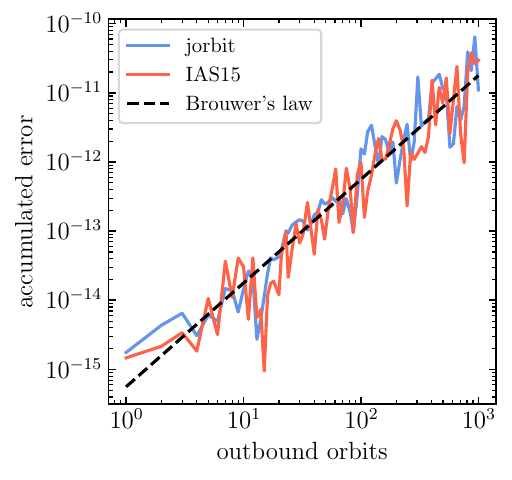}
    \caption{A comparison of accumulated errors between \jorbit and IAS15 for a simple scale-free $r=1, P=2\pi$ circular orbit of a tracer particle around a unit mass. Both integrators were run forward to the specified number of outbound orbits, then reversed back to zero. The absolute difference between the start and finish positions gives a measure of the error accumulated purely due to the numerical method. Here we have plotted half this distance to show the degree of error expected for an integration of a given length. Note \jorbit stays well below its $10^{-10}$ target for many hundreds of dynamical times, meaning numerical imprecision should never be the limiting factor in an orbit fit to real data. Brouwer's Law, where error $\propto t^{3/2}$, is included for reference. \pycodelink{https://github.com/ben-cassese/jorbit/blob/907e5b8ccee42479bf4fa67435b5733318d0bd9d/paper/out_and_back_tests.ipynb}}
    \label{fig:ias15_rebound_compare}
\end{figure}

The IAS15 integration algorithm, first outlined in \citet{everhart_1985} and improved in \citet{rein_ias15_2015}, meets this steep precision requirement by converging each step to near machine-precision before proceeding. Through its use of compensated summation and careful ordering of arithmetic operations, IAS15 achieves ``optimal'' numerical behavior where the errors (including those impossible to suppress due to round-off and floating point truncation) grow according to Brouwer’s Law, or $\propto t^{3/2}$ (see Fig. \ref{fig:ias15_rebound_compare}).

An open source implementation of IAS15 makes up the core integration routine of the \rebound ecosystem. Unfortunately, we could not leverage this well-documented and supported library since \rebound is written in the \texttt{C} programming language which, while extremely performant, does not support autodifferentiation and is challenging to bind to \jax. Consequently, we created a re-implementation of IAS15 entirely in \jax.

In the end, we produced two separate just-in-time compiled, autodifferentiable integrators based on IAS15. The first is modeled as closely as possible to the \rebound source code, matching arithmetic operations like-for-like wherever possible. Due to a) unintentional implementation differences and b) difference between the \texttt{C} and \jax compilers, the final product of this effort does not match \rebound's behavior to within machine precision. It does, however, qualitatively match \rebound's behavior on timescales relevant to orbit fitting, as seen in Fig. \ref{fig:ias15_rebound_compare}. We use the time stepping criteria from \citet{pham_timestep_2024}, which in early 2024 became the default in \rebound.

Our other integrator is a generalized adaptation of the IAS15 algorithm and differs in four key ways. Firstly, it uses a custom ``DoubleDouble'' \jax object that follows the arithmetic rules for extended-precision operations in \citet{dekker_1971}. By representing every value as two floating point values instead of one, this allows for simulations at a numerical precision that exceeds the limitations set by individual floating point operations. This could be useful either for testing other routines, actually running high-precision simulations, or enabling \jorbit simulations in hardware-limited scenarios. We note that certain CPU architectures naturally support 80 or 128 bit floating point values, and that the INPOP series of published ephemerides harnesses these extended-precision capabilities (e.g., \citet{inpop06}). However, this capability is not included in all common CPUs, and to run simulations that accurately resolve a wide range of perturber masses, users may need to use software to mimic these extra bits of precision. Additionally, many GPUs only support 32 bit operations: while there are not many scenarios where GPUs would substantially improve \jorbit's performance\footnote{For single-particle orbit fits, most of the computation time is spent on necessarily sequential operations as the integrator steps through time. There are certainly advantages of GPU acceleration for larger simulations with many ($>1000$) particles, though \jorbit's default integrator settings were not set with these in mind.}, any user taking advantage of \jax's ability to cross-compile to different hardware should be aware of the precision limitations of their accelerators.

\begin{figure}
    \centering
    \includegraphics[width=\columnwidth]{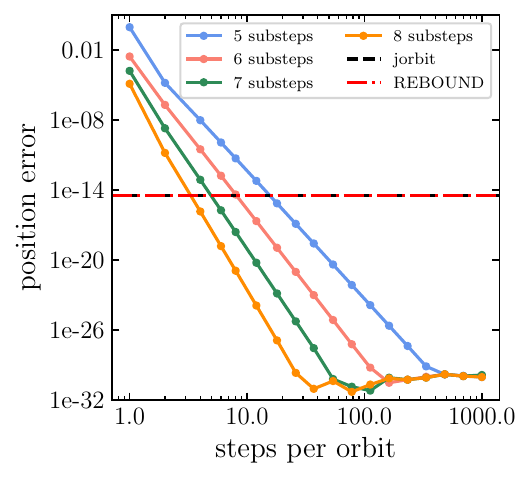}
    \caption{Example performance of the ``DoubleDouble'' precision integrator. Note we can tune the order of the integrator based on the number of substeps, and that we can achieve relative position errors near roughly $10^{-32}$, about 16 orders of magnitude below the best precision available to double-precision operations. Were we using single-precision operations, like those available on most GPUs, the DoubleDouble operations would level out at $10^{-16}$. The single-orbit error from \rebound's implementation of IAS15 and \jorbit, both of which use 7 substeps, are shown as a horizontal lines since both use adaptive, rather than fixed, time steps. \pycodelink{https://github.com/ben-cassese/jorbit/blob/907e5b8ccee42479bf4fa67435b5733318d0bd9d/paper/doubledouble_integrator.ipynb}}
    \label{fig:iasnn_orders}
\end{figure}

Secondly, IAS15 achieves its 15th-order error suppression behavior by evaluating the acceleration function at 7 ``substeps'' for every integration step, where the substeps are placed according to Gauss-Radau spacings in time. This version of the algorithm similarly uses Gauss-Radau spacings, but allows the user to choose the number of internal substeps, and consequently, the order of the integrator (see Fig. \ref{fig:iasnn_orders}). The required coefficients are computed once before each run using the arbitrary-precision python package \texttt{mpmath} \citep{mpmath}, and are stored as \jax-compatible arrays prior to any inference-related computations. We verified that the resulting constants match those listed in Table 12 of \citet{gaussian_quad_book}.

Thirdly, these DoubleDouble-precision routines use a fixed, rather than adaptive time step. This is done to minimize the amount of logic-based control flow which can significantly slow down GPU applications. This is a limitation we accept for the default integrator, since IAS15 must make several run time decisions about how many times to execute certain parts of the algorithm. For example, each time step relies on a predictor-corrector scheme that iterates until convergence. On a larger scale, since IAS15 determines the most appropriate step size, we cannot know a priori how many steps will be required to reach a certain target time. These run time decisions do not play well with \jax's JIT compiler, which in general prefers to know the exact computation graph at compile time.

Though we experimented with fixed-iteration implementations with generous padding/convergence checking, we found that the memory requirements of these approaches hobbled their performance. Acknowledging this, we chose to use ``while-loop'' like structures in the default algorithm, which precludes \jax's ability to use reverse-mode autodifferentiation. However, forward-mode is still available, and consequently whenever \jorbit computes a derivative that involves passing through its default numerical integrator, it does so via forward-mode autodifferentiation.

Fourth and finally, the DoubleDouble-precision routines are significantly slower than the default standard IAS15 implementation. This is unsurprising since the algorithms of \citet{dekker_1971} mandate that every arithmetic operation done at standard precision must now be represented via many individual operations. This is why, despite the superior error suppression of the DoubleDouble routines, we leave \jorbit's IAS15 implementation as its default integrator.

\subsection{Acceleration Building Blocks} \label{sub:accelerations}

As mentioned in Sec \ref{sub:philosophy}, \jorbit is capable of integrating nearly arbitrary acceleration functions, but has two fundamental ones built in by default: Newtonian gravitation and Parameterized Post-Newtonian (PPN) gravitation.

The former is straightforward enough: we simply evaluate Newton's law for every pairwise interaction. \jorbit does take care, however, to avoid unnecessary interactions between massless particles. We have validated that our acceleration evaluations agree with those predicted by \rebound across a wide range of simulation setups \pycodelink{https://github.com/ben-cassese/jorbit/blob/907e5b8ccee42479bf4fa67435b5733318d0bd9d/tests/test_accelerations.py\#L67}.

To incorporate general relativity (GR) effects, we implemented a Parameterized Post-Newtonian scheme that closely mimics the \texttt{gr\_full} routine in \texttt{REBOUNDx} \citep{reboundx}, which itself is based on the algorithms from \citet{newhall_1983}. This is a conservative choice that trades extra computation for the consideration of every massive body's contribution to the GR-corrected accelerations. Though the solar system is dominated by a central mass (the sun), an arrangement that suggests that a more approximate scheme such as that presented in \citet{nobili_1986} might suffice, we found that the extra precision afforded during closer interactions with Jupiter was worth the extra operations.

This scheme relies on iterations until convergence is reached, just like parts of the IAS15 integrator described in Sec \ref{sub:integrator}. We have verified that our accelerations agree with those predicted by \texttt{REBOUNDx} to within machine precision \pycodelink{https://github.com/ben-cassese/jorbit/blob/907e5b8ccee42479bf4fa67435b5733318d0bd9d/tests/test_accelerations.py\#L16}.

In addition to these gravitational effects, we have also included the ability to add contributions from non-gravitational accelerations through the \citet{marsden_1973} parameterization, and the ability to assign arbitrary-order gravitational harmonics $J$ to any body included in the DE ephemeris. Sec. \ref{sub:apophis} and Fig. \ref{fig:apophis} demonstrate a simulation using these additional forces, though they are not included by default  when using the \texttt{Particle} interface.

\subsection{Including the JPL DE440 Ephemeris} \label{sub:jpl_ephem}

In its default operating mode, \jorbit will compute the gravitational acceleration on particle(s) due to the solar system planets\footnote{For \jorbit's purposes, the ``planets'' are all of the bodies in the \href{https://ssd.jpl.nasa.gov//ftp/eph/planets/bsp/}{de440.bsp} file: the Sun, Mercury, Venus, Earth, Moon, Mars, Jupiter Barycenter, Saturn Barycenter, Uranus Barycenter, Neptune Barycenter, and Pluto Barycenter.} (via PPN gravity) and the 16 most massive asteroids\footnote{The asteroids in the \href{https://ssd.jpl.nasa.gov//ftp/eph/small_bodies/asteroids_de441/}{sb441-n16.bsp} file: (1) Ceres, (2) Pallas, (3) Juno, (4) Vesta, (7) Iris, (10) Hygiea, (15) Eunomia, (16) Psyche, (31) Euphrosyne, (52) Europa, (65) Cybele, (87) Sylvia, (88) Thisbe, (107) Camilla, (511) Davida, and (704) Interamnia} (via Newtonian gravity). However, it does not actually integrate all of these perturbers alongside the particles of interest: instead, at every (sub)time step, it uses the JPL DE440 ephemeris \citep{park_2021} to look up the positions and velocities of these objects, uses these as inputs to the particle's acceleration, then discards them before the next time step.

Technically this can lead to simulations which are not self-consistent: if our particles have non-zero mass, or if our acceleration functions differ from those used in the creation of DE440, our perturbers will not follow the trajectories suggested by our simulation. In practice, however, the small logical tension is a more than fair price to pay to a) avoid integrating many more particles and b) maintain a close agreement with the Horizons service, which we risk losing if we let the planets run rogue every simulation.

To query the time-dependent states of the perturbers, we first use \texttt{astropy} to automatically download and cache the relevant DE440 ephemeris .bsp files. These store coefficients that describe the $x,y,z$ positions of each perturber as piecewise Chebyshev polynomials. We assume the velocity of each particle is given by the instantaneous derivative of its position polynomial, similar to \assist. Although these files are $\sim$800 MB total, by trimming them to only include a limited time span, we can avoid loading a large array into memory (e.g., loading just the coefficients describing the planets between 1980-01-01 and 2050-01-01 requires < 15 MB). This extraction stage relies on the \texttt{jplephem} library (which also underpins the \texttt{astropy} ephemeris ecosystem, \citet{jplephem}), though once arrays are loaded, everything that follows involves custom \jax implementations.

We next pad/join these coefficients into a single array which allows us to vectorize over all particles instead of looping over each perturber individually. Though exact benchmarks will vary across systems, in our development environment, solving for the position/velocity of the 11 ``planets'' in DE440 at a specific time takes $<60 \mu$s \pycodelink{https://github.com/ben-cassese/jorbit/blob/907e5b8ccee42479bf4fa67435b5733318d0bd9d/paper/ephemeris_timing.ipynb}.

We note that there are other tools capable of processing the JPL DE ephemeris files, including SPICE \citep{acton_1996} and its python interface, \texttt{SpiceyPy} \citep{annex_2020}, \texttt{CalcEph} by the IMCEE\footnote{\href{https://www.imcce.fr/recherche/equipes/asd/calceph/}{CalcEph}}, \texttt{FIRE} \citep{arora_2010}, and \texttt{Ephemerides.jl}\footnote{\href{https://github.com/JuliaSpaceMissionDesign/Ephemerides.jl}{Ephemerides.jl}}. The last of these is also capable of autodifferentiation. However, \jorbit's re-implementation of their capabilities in \jax was crucial to maintain its end-to-end \jax-only workflow.

\subsection{Converting to sky coordinates} \label{sub:skycoord}
Though up to now we have only considered 3D barycentric positions, to compare its integrations to actual observables \jorbit must be able to convert from 3D locations to 2D on-sky astrometry as seen from a specific observer. This is not necessarily a straightforward computation, since both light travel time delays and general relativistic deflections can affect the final result. Following JPL Horizons, we correct for the former, but not the latter.

We implemented a JAX-routine that closely mirrors \texttt{astropy}'s \texttt{\_get\_apparent\_body\_position} function within the \texttt{coordinates.solar\_system} module. This routine ``guesses'' a light travel time correction, integrates the particle in question to that corrected epoch, and re-evaluates the distance between the observer and the particle. It then updates its guess and iterates a fixed number of times to refine its estimate of the final delay. For every guess/small integration iteration, \jorbit uses the same integrator/acceleration function that was used to compute the original barycentric position. Even though these corrections are often quite small ($\sim$minutes for main belt asteroids), we found that relying on a high-order integrator and including all of the massive perturbers led to the best agreement with predictions from JPL Horizons.

\section{A Local MPChecker} \label{sec:mpchecker}

Here we pivot away from orbit-fitting related functionality and to the second wing of \jorbit's design: its implementation of ``MPChecker''-like functions. 

As described in Sec \ref{sec:intro}, although computing the on-sky ephemeris of a given minor planet as seen from a given location is fairly straightforward, the related problem of identifying the minor planets in a given patch of sky at a given time from a given observatory is much more challenging.

If users have actual data and have extracted astrometry of all the minor planets, the MPC recommends submitting these raw detections and allowing them to match them to the known objects. Our use case is slightly more general, however, since we consider scenarios in which the user has either not resolved the individual objects, or has not collected data in the first place (and is potentially considering some time in the future).

\subsection{Initial Simulation/Caching} \label{sub:big_sim}

A simple procedure for checking which minor planets fell near a certain coordinate at a certain time could proceed as follows: one could store the initial conditions of every known object at a certain epoch, integrate them forward to the desired time, compute their observatory-specific sky positions (which must include a per-particle light travel time correction, which itself requires per-particle integration to a specific time), then sort based on angular distance to the desired location. While this approach could work, it is likely too slow/computationally demanding to be practically realized.

\jorbit instead takes a two-step approach: we did indeed simulate the motions of a large number of particles, but we then compressed and cached these simulations as a starting place for \jorbit's user-facing functions. More specifically, we retrieved the Minor Planet Center's MPCORB.DAT file in February 2025 and extracted the names of all 1,438,635 then-known numbered and provisionally designated minor planets.\footnote{Note: MPCORB.DAT excludes comets and comet fragments: as a consequence, \jorbit will fail to identify cometary materials.} We then queried JPL Horizons for the barycentric positions of each of these objects on 2020-01-01\footnote{Six objects, or 0.0004\% of the total non-comets, did not have Horizons states available on that date and were excluded. In their packed designation form, they are: K19M00O, K18L00A, K08T03C, K24Y55D, K10DB6X, and K14A00A.}.

Next, we integrated the motion of each object forwards in time to 2040-01-01 and backwards to 2000-01-01, logging each 3D position every 10 hours during that span. These integrations did take perturbations from the planets/large asteroids into account, as usual, but only considered Newtonian gravity since the following step would have rendered the extra precision of PPN useless anyway\footnote{Also, due to the timing of our development process, the 2020-2040 simulations used the Earth-Moon barycenter as a perturber, while the 2000-2020 simulations resolve the Earth and the Moon as individual perturbers. This is the current \jorbit default.}. Finally, we converted these 3D positions to on-sky positions as seen by an observer at the geocenter. This introduces a $\sim$tenths of an arcsec offset with observers located on the surface, potentially 10s-100s of arcsec with observers based in Earth orbit, and even larger discrepancies when considering assets at L2. These offsets are all much larger than the typical relativistic corrections.

Instead of saving all of these sky coordinates in raw form, we compressed them into two 11th order piecewise Chebyshev polynomials (one for RA, one for Dec) broken into 30 day chunks. The polynomial order, temporal coordinate density, and chunk size were all chosen heuristically and could be targets for future improvements. Each 30-day chunk was then saved as an individual $\sim$250 MB numpy array, and all files were uploaded to the HuggingFace repository\footnote{\jorbit's dataset can be found \href{https://huggingface.co/datasets/jorbit/jorbit_mpchecker/tree/main}{here}.} of public machine learning datasets. The total amount of cached data is roughly 120 GB, though as described below, users only need to locally download the time chunks relevant to their investigation.

We note that this procedure very closely mimics ones used for creating a DE-like ephemeris, which similarly produces piecewise Chebyshev polynomials. Our cached material, however, represent sky positions, whereas most ephemeris files cache 3D positions. Half of these simulations were carried out on the Grace high-performance computing cluster at Yale University, while the other half were run on the Anvil cluster \citep{anvil} at Purdue University with an allocation granted via the NSF ACCESS program \citep{nsf_access}.

\subsection{A User's Workflow} \label{sub:mpchecker_workflow}

\jorbit maintains a simple interface for users wanting to know the identity and properties of minor planets near a certain region. All of the downloading/caching happens automatically and is built on top of the \texttt{astropy} data cache.

What exactly happens under the hood depends on the level of inaccuracy a user is willing to tolerate. If they are willing to accept the errors described above that were introduced by the use of Newtonian gravity and a geocentric observer, then \jorbit's job is fairly simple: it downloads the relevant ephemeris chunk, evaluates the piecewise polynomials at the desired time (which it can do rapidly via vectorized operations, as evaluating polynomials is much faster than numerical integration), then returns all particles that fell within the requested radius.

If the user either a) wants higher-precision astrometry or b) wants the real-time apparent magnitude of each object (information which requires knowing its 3D position, which we lost by only saving a compressed form of sky coordinates), then \jorbit must do more work. The process is as follows:

\begin{enumerate}
    \item It runs the same geocentric-based computations as a low-resolution checker to flag all asteroids that fall within the desired radius. Note that any particles which fell outside this radius due to the initial inaccuracies could be missed, so users are encouraged to inflate their desired radii.
    \item Next, it relies on its internally cached initial state of those objects and runs a full N-body integration to propagate them forward to the requested time. By default, this includes gravitational perturbations from all of the usual 27 bodies from DE 440.
    \item It then queries JPL Horizons for the barycentric position of the observer at that time (these can be cached if one wishes to check multiple coordinates observed from the same observatory at the same times) and converts the 3D particle positions into on-sky coordinates. This is the only part of the procedure that relies on an external service.
    \item Finally, it computes each particle's time-dependent phase angle and distance, combines these with estimates of each particle's absolute $H$ magnitude stored in the MPCORB.DAT file, and computes a time-dependent, observer-specific apparent magnitude for each.
    
\end{enumerate}

Again, benchmarks will vary widely across different systems and situations (it takes longer to simulate more crowded regions of the sky), but, to give a sense of scale in our development environment, evaluating the low-resolution version at a single timestamp takes $<1$ second, and a high-resolution query across 1.5 days sampled every 15 minutes takes 25 seconds. To speed up operations, when a user requests a time series around a single point, \jorbit first computes the low-resolution position of every object at the midpoint of that time series, then moving forwards only considers particles that fell within 30 degrees of the requested position. Again, this value was chosen heuristically and could be changed in future developments, but for now, users interested in rapidly-moving NEOs over moderate time spans for a single location should be aware of this choice.

In an era of 100s of new minor planets discovered each month and at the dawn of an era sure to be dominated by discoveries from the Vera C. Rubin Observatory \citep{lsst}, we expect to need to update the cache \jorbit simulations regularly. Our tentative plan is for bi-annual releases and to eventually implement an $H$ magnitude cutoff, though these choices are subject to change. By default \jorbit will automatically re-download any mpchecker-related files that were initially downloaded more than six months ago. We stress that we do not expect MPChecker results to be consistent across version releases since they will be built from potentially different initial state vectors.

\section{Examples and Applications} \label{sec:examples}

\subsection{Comparison with JPL Horizons} \label{sub:comparison}

Though in Fig. \ref{fig:ias15_rebound_compare} we demonstrated that our underlying integrator performs similarly to \rebound's default integrator in simple circumstances, in the context of orbit fitting, the precision of the 3D position estimate is not what counts: when dealing with astrometry, the only quantity we can actually observe is the position on the plane of the sky.

\begin{figure*}
    \centering
    \includegraphics[width=0.8\textwidth]{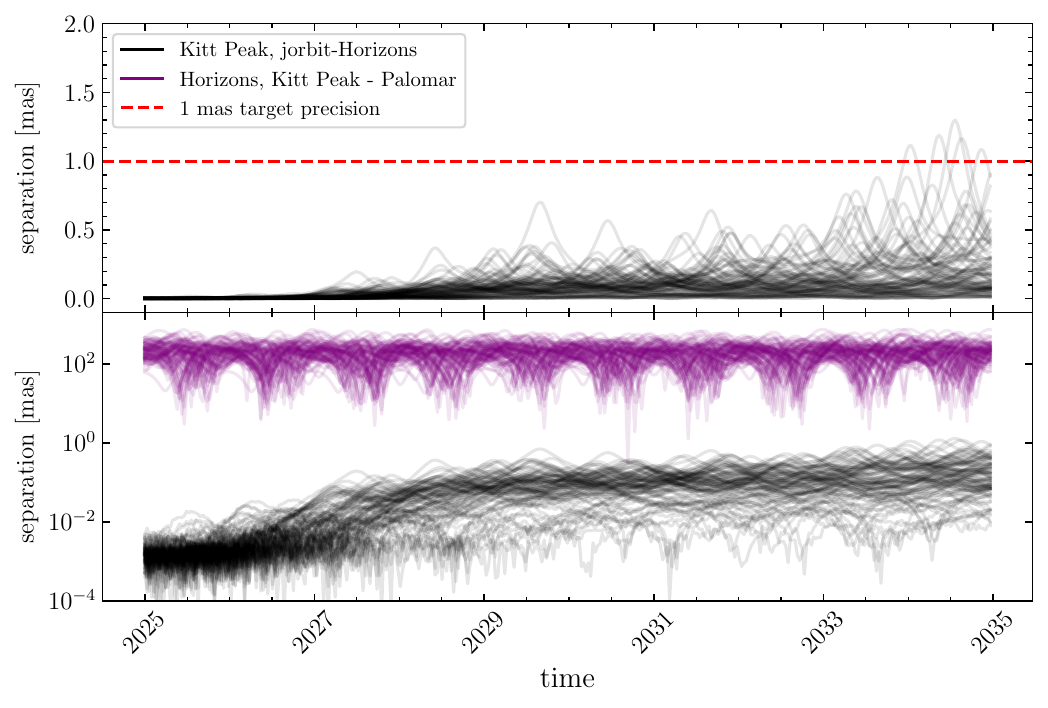}
    \caption{Top: The difference in predicted sky positions for 100 randomly sampled objects as calculated by JPL Horizons and \jorbit. Bottom: The same data on a log scale alongside the difference in sky positions of the same objects at the same times, as viewed from Kitt Peak and Palomar Mountain and predicted by Horizons. After 10 years of integration, the median disagreement between Horizons and \jorbit was 0.13\,mas, while the median effect of the parallax between the observatories was 209\,mas. \pycodelink{https://github.com/ben-cassese/jorbit/blob/907e5b8ccee42479bf4fa67435b5733318d0bd9d/paper/horizons_compare.ipynb}}
    \label{fig:horizons_compare}
\end{figure*}

Accordingly, a more informative test is to compare \jorbit's predictions of minor planet sky positions with those from an external service like JPL Horizons. We do this in Fig. \ref{fig:horizons_compare}: we randomly sampled 100 numbered minor planets by drawing random integers, queried Horizons for their state vectors at 2025-01-01, then used these as initial conditions for \jorbit integrations. We integrate each particle forward 10 years, logging its sky position as seen from Kitt Peak National Observatory every 10 days. We then separately query Horizons for its prediction for the astrometric coordinates seen from the same location at the same time, and plot the differences. Note that this procedure includes many days where the objects would not actually be visible due to proximity to the sun, and that during these times the angular differences between methods will be suppressed due to the larger distance to the targets. Thus, this is a somewhat unfairly optimistic measure; however, we accept the procedure for the sake of keeping sampling simple and because even the worst-case disagreements at each time step still demonstrate our success at mimicking Horizons.

The agreement between the two predictions is quite good, remaining near the $\mu$-arcsec level for several years before diverging, though remaining just a fraction of a mas for most objects for the duration of the simulation. This implies that our underlying implementation of both the integration engine and our accounting of perturbations from massive planets/asteroids aligns nicely with Horizons'. To provide a sense of scale we also queried Horizons for the astrometric positions of each of these objects as seen from Palomar Observatory, a facility just $\sim500$\,km from Kitt Peak, at the same times as the original simulated observations. By the end of 2034, the median disagreement between Horizons and \jorbit was 0.13\,mas, while the median effect of the tiny parallax between these observatories was 209\,mas.

\subsection{Main-belt Asteroid Fits} \label{sub:wiki_fits}

We can use \jorbit to forward model a minor planet ephemeris given a set of model parameters. These parameters can be essentially anything related to minor planet propagation and astrometry: for instance, we could tweak the precession/nutation rates of the Earth, or the mass of Saturn, or the $A$ coefficients describing non-gravitational acceleration in the \citet{marsden_1973} convention. So long as the cumulative acceleration model can be written in \jax, \jorbit can create a resulting ephemeris, compute the likelihood of that particular parameter vector, and take the derivative of the likelihood with respect to each parameter.

The most common model we anticipate users fitting is a simple orbit: six parameters associated with a specific epoch that describe a particle's state and nothing else. These can be expressed in any form that can be reduced to a barycentric Cartesian position and velocity, including as osculating Keplerian elements. By default within its \texttt{Particle} class, \jorbit will evolve this state vector under the influence of the sun and 11 perturbers via PPN gravity (the ``planets'' listed earlier) and 16 perturbers via Newtonian gravity (the asteroids). We include several convenience methods for fitting \texttt{Particle}s, including \texttt{residuals}, which computes the sky-projected astrometric residuals implied by a given state, and \texttt{max\_likelihood}, which uses the \texttt{scipy} \citep{scipy} implementation of the L-BFGS-B algorithm to maximize the log-likelihood. This also uses the exact gradients from \jorbit's automatic differentiation instead of finite differences. For a comparison to derivatives as computed via autodiff versus those computed via finite differencing, see Appendix \ref{app:autodiff} and Fig. \ref{fig:autodiff_works}, which demonstrate excellent agreement between the two.

Note that when we discuss ``likelihood'', we implicitly assume that the astrometric errors can be modeled as a 2D Gaussian in the tangent plane of the sky. \jorbit can handle arbitrary covariance between RA and Dec directions (e.g., we can faithfully model \textit{Gaia}'s extremely unequal uncertainties along vs. across its scanning direction; \citet{spoto_2018}), but it assumes that errors are small enough that fuller treatments of spherical geometry, like the Kent distribution \citep{kent_1982}, are not necessary. \jorbit also rotates all observations to the equator before computing the tangent plane projection to handle complications arising near the poles.

\begin{figure*}
    \centering
    \includegraphics[width=\textwidth]{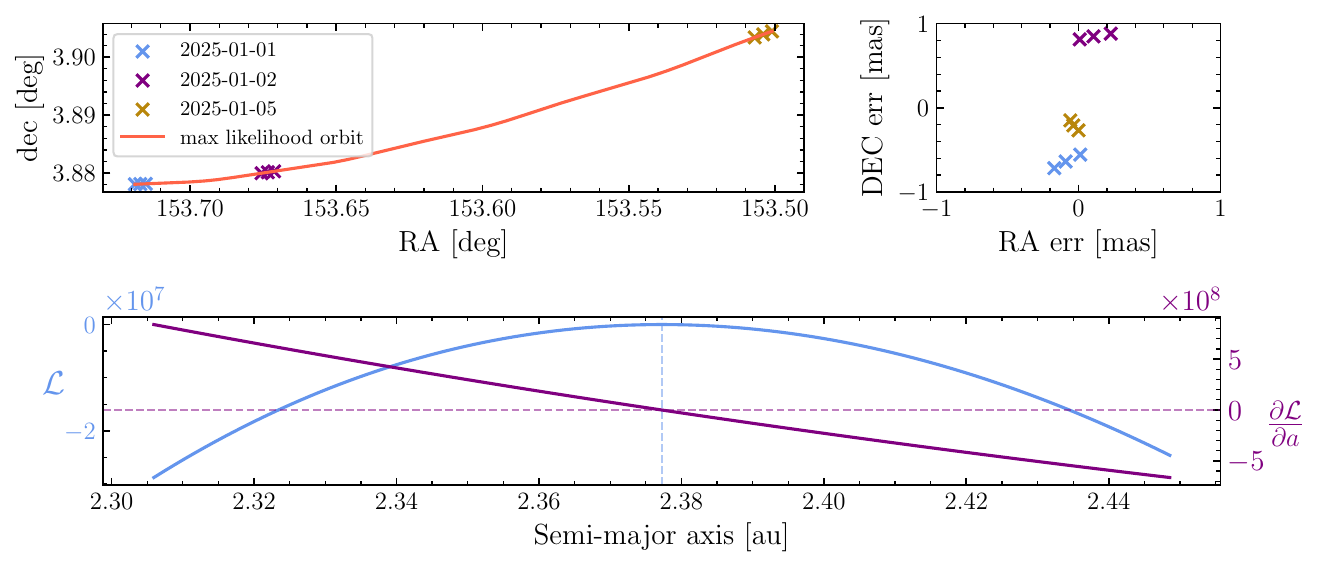}
    \caption{Simulated observations of and \jorbit's prediction for the maximum likelihood orbit of (274301) Wikipedia. To demonstrate \jorbit's ability to compute the likelihood of a set of parameters and the derivatives of that likelihood, we show the results of holding all other best-fit parameters fixed while varying the semimajor axis. The 1st and 2nd derivatives were computed via automatic differentiation of the fully dynamical model (i.e., it was propagated through the IAS15-integrated, perturbers-included, GR-included N-body simulation), not through finite differencing or manual algebra. To guide the eye, the vertical line marks the semimajor axis of the fit, while the horizontal line marks zero derivative of the likelihood with respect to semimajor axis. \pycodelink{https://github.com/ben-cassese/jorbit/blob/fe75215cce60a6f604532a0c645eee94c77bd6b8/paper/wikipedia_orbit_fit.ipynb}}
    \label{fig:max_likelihood_orbit}
\end{figure*}

In Fig \ref{fig:max_likelihood_orbit}, we show the results of an orbit ``fit'' (just a maximum likelihood estimate, not a full joint posterior of each component of the state vector) to simulated observations of the minor planet (274301) Wikipedia. We queried Horizons for nine astrometric positions of this object spread across three nights, with three observations per night separated by one hour. The first two nights are consecutive, while the third is three days later. This dataset is meant to represent a discovery scenario with a limited baseline. We intentionally did not add noise to these observations to assess \jorbit's ability to recover the true injected state vector.

Under the hood, if at least three observations are available, \jorbit first uses a \jax implementation of Gauss's method for orbit fitting to derive a reasonable starting point for its computations. It then proceeds with the L-BFGS-B algorithm until convergence is reached, and returns the resulting set of osculating Keplerian elements. In this case, given these nine observations and no other information about the particle, \jorbit was able to find a solution with sub-mas residuals and osculating elements all within 0.4\% of those assumed by Horizons \pycodelink{https://github.com/ben-cassese/jorbit/blob/907e5b8ccee42479bf4fa67435b5733318d0bd9d/paper/wikipedia_orbit_fit.ipynb}.

We intend to further develop \jorbit's fitting abilities in the future to include methods that compute not just a maximum likelihood solution, but also the full posterior of all model parameters. This could come in the form of Hamiltonian Monte Carlo routines or other algorithms that leverage information about the curvature of the posterior.

\subsection{Unknown Asteroid Identification} \label{sub:unknown_asteroid}

Next we imagine a scenario in which we have observed a certain patch of sky at a certain time, have found a moving object, and want to identify it. We will use the unsettlingly specific choices of RA=06:29:26.9, Dec=+15:39:20.3, Time=2023-11-23 00:00 for reasons that will become clear in Sec. \ref{sub:light_curve_contamination}.

Though the exact code API may evolve in the future, here we demonstrate the current high-level wrapper around all of \jorbit's \texttt{mpchecker}-related functions:

\begin{lstlisting}
import astropy.units as u
from astropy.coordinates import SkyCoord
from astropy.time import Time
from jorbit.mpchecker import mpchecker

mpchecker(
    coordinate=SkyCoord(97.36206*u.deg, 15.65564*u.deg),
    time=Time("2023-11-23"),
    radius=5*u.arcmin,
    extra_precision=True,
    observer="Kitt Peak",
)
\end{lstlisting}

This query produces an \texttt{astropy} table with 2 rows, one for each minor planet that fell within $5'$ of our target coordinate at the time of observation \textit{as seen from the geocenter}. We emphasize that while each actual entry in the table contains information that takes the specific observatory location into account (e.g., its astrometric coordinate and estimated $V$-band magnitude), the choice of \textit{which} objects to include in the table was based on the geocenter-based simulation. For ground-based observers, the impact of this slight parallax will be small, but for space-based observers, especially those operating beyond low Earth orbit, it could be several arcmin.

For this query, both \jorbit and the MPC's online MPChecker service predict that the objects (1055) Tynka and (172962) 2005 MZ13 fall within our radius of interest. We compare the properties of the two predictions in Table \ref{tab:mpchecker}: note that, aside from querying Horizons for the barycentric coordinates of the observer at this particular time, the rest of \jorbit's computations were done entirely offline.

\begin{deluxetable*}{lcccr}
\label{tab:mpchecker}
\tablecaption{
\jorbit's and the MPC's predictions for which known minor planets appear within 3$\arcmin$ of the target star at the requested epoch, together with their coordinates and magnitudes. Note that the MPC also predicted that 2015 XB347 fell within a 3$\arcmin$ radius; see text for details. Quantities are ordered by (their value as calculated offline by \jorbit)/(their value as calculated online by the MPC). \pycodelink{https://github.com/ben-cassese/jorbit/blob/907e5b8ccee42479bf4fa67435b5733318d0bd9d/paper/asteroid_identification.ipynb}}
\tablewidth{0pt}
\tablehead{
\colhead{Object Name} & \colhead{R.A} & \colhead{Dec} & \colhead{Visual Magnitude}
}
\startdata
(1055) Tynka & 06:29:33.93/06:29:34.0 & +15:40:45.75/+15:40:46 & 15.49/15.5 \\
(172962) 2005 MZ13 & 06:29:18.98/06:29:19.0 & +15:42:55.72/+15:42:56 & 20.51/20.5 \\
\enddata
\end{deluxetable*}

We again note that \jorbit's MPChecker functions will fail to identify cometary material: this is because comets and their fragments do not appear in the MPCORB.DAT file, and although \jorbit has built-in functions to handle non-gravitational acceleration, it does not have access to the object-specific coefficients required to use it without additional user input. However, since minor planets make up the bulk of known solar system objects ($>1.4$ million vs. $<5000$, according to the MPC as of writing), the majority of queries run during development revealed no disagreements, though it is difficult to test the agreement between \jorbit and the MPC's MPChecker service at scale due to the required computation time on the MPC side.

\subsection{Light Curve Contamination} \label{sub:light_curve_contamination}

While the ability to identify minor planets near a given coordinate at one specific instance is helpful for unknown object identification, \jorbit's ability to run these calculations locally truly shows its benefits when we string many of these queries together into time-dependent simulations. This allows us to examine how all known solar system objects affect a time series observation, which can be useful even in scenarios in which we do not resolve the individual objects. In what follows we describe a science case involving a search for transiting exoplanets, but similar arguments apply for any investigations involving time series measurements near the ecliptic. We plan to explore this type of investigation at scale in a companion paper and here simply demonstrate that \jorbit can aid modelers wishing to check/correct their time series observations for contamination from minor planets.

\begin{figure}
    \begin{interactive}{animation}{nearest_asteroid.mp4}
    \centering
    \includegraphics[width=0.9\columnwidth]{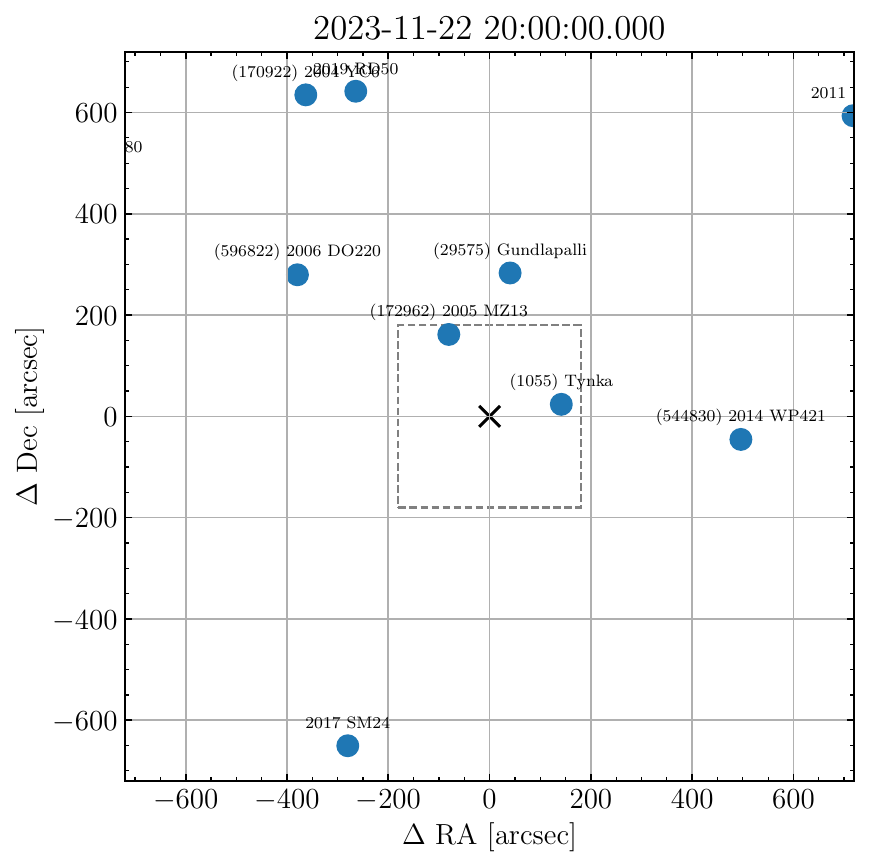}
    \end{interactive}
    \caption{A freeze-frame of a \jorbit-created animation of minor planets moving near a target star of interest in the TESS dataset. An animated version of the figure may be found in the online version of this article. Time-resolved summary statistics from this simulation are shown in Fig. \ref{fig:lightcurve_contamination}. The dashed rectangle has a diameter of 3.85\arcmin, which is the side length of the Target Pixel File (postage stamp) for this target.\pycodelink{https://github.com/ben-cassese/jorbit/blob/907e5b8ccee42479bf4fa67435b5733318d0bd9d/docs/tutorials/lightcurve_contamination.ipynb}. The animation runs from Nov. 22nd, 2023 at 12:00:00 to Nov. 23rd 11:57:00. The real-time duration is 16 seconds.}
    \label{fig:nearest_asteroid_frame}
\end{figure}

Ground- and spaced-based photometric surveys designed to detect transiting exoplanets often lock onto one region of the sky and take repeated images of many stars at once. Several of these surveys, despite their explicit goal of imaging targets beyond the solar system, routinely observe the ecliptic plane. These most notably include the K2 mission \citep{k2}, which was forced to observe in specific regions following mechanical failures aboard the \textit{Kepler} spacecraft, and the Transiting Exoplanet Survey Satellite (TESS, \citet{tess}), which occasionally rotates its field of view to observe targets within the ecliptic, in part to follow up on K2 discoveries.

Though these surveys are ``shallow'' in the sense that their single-frame detection limits are quite bright (TESS, for example, has only a 10\,cm aperture for each of its four cameras), they rely on and are capable of exquisitely precise differential photometry to detect transiting planets. As an example, a Jupiter-sized planet transiting a Sun-sized star causes a change of flux of approximately 1\%. This implies that a foreground object 5 magnitudes fainter than the primary target could, should it wander into then out of the extraction aperture, complicate a search for more distant planets. Considering that about 37\% of all TESS targets selected for high-cadence 20\,s observations in Sectors 42-46, 70-72 (the ecliptic plane sectors pre-2025) have $T_{\rm mag}>15$, a sizable fraction of these targets could be vulnerable to contamination from $T_{\rm mag}\sim20$ minor planets.

Critically, these minor planets do not have to approach particularly close to the target star to impact the observations. The SPOC TESS processing pipeline \citep{tess_spoc}, for example, compares the total flux contained within several 21$\arcsec$ pixels centered on the star with several pixels further away to subtract the sky background and correct for instrumental effects. If a coincident minor planet appears in the extraction aperture, the star will appear temporarily brighter; if it appears in the background aperture, however, the star will appear temporarily fainter by comparison.

Let us now consider TIC 438412198, a star selected as a 20\,s high-cadence target for TESS by general investigator program G06144 (Hord, Cycle 6). This star sits at the coordinates previously used in Sec \ref{sub:unknown_asteroid}, has $T_{\rm mag}=10.08$, and was observed from November 11 through December 7, 2023 in Sector 72. 

\begin{figure*}
    \centering
    \includegraphics[width=0.8\textwidth]{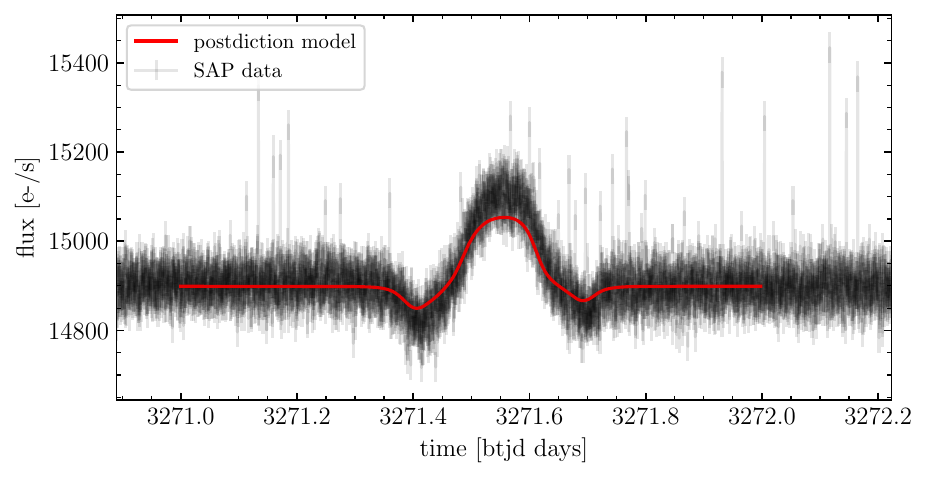}
    \caption{Actual TESS data of TIC 438412198 and \jorbit's postdiction simulation of the total flux of minor planets, centered on the flyby of asteroid (1055) Tynka. Note that transient asteroids can produce both apparent dips and apparent spikes in Simple Aperture Photometry light curves. The initial dimming occurs when (1055) Tynka approaches a cluster of pixels in the ``background'' aperture used to determine the sky background but before it reached the pixels composing the extraction aperture. The exact morphology of each minor planet/TIC interaction depends on the relative placement of the background aperture, extraction aperture, and the minor planet's trajectory. \pycodelink{https://github.com/ben-cassese/jorbit/blob/907e5b8ccee42479bf4fa67435b5733318d0bd9d/paper/lightcurve_contamination.ipynb}}
    \label{fig:lightcurve_contamination}
\end{figure*}

We used \jorbit's MPChecker functions to investigate the impact of minor planets that approached near the star during TESS's observations. We found that during the 24-hour period beginning on 12:00 November 23rd, two minor planets passed within 3\arcmin\, of the target star, close enough to cross the target's ``postage stamp'', or small portion of the TESS CCD that is read out every 20\,s. \jorbit post-dicted that the brighter of these, the asteroid (1055) Tynka, had a $V$\,magnitude of 15.5 at the time of the encounter, while the other body, (172962) 2005 MZ13, had $V=20.5$. Using \jorbit's post-diction of the minor planets' trajectories and magnitudes, we can fully model the impact of this transient on the recorded Simple Aperture Photometry (SAP) light curve.

To create this model, we first used the World Coordinate System information contained within the postage stamp header to convert the on-sky trajectory of each asteroid to an on-detector trajectory. Next, using the known model of TESS's Pixel Response Function, we created a mock image for each time stamp that included the two asteroids as point sources at their appropriate positions/time-resolved $V$ magnitudes. We then converted from $V$\,mag to TESS's $T$ band using mean asteroid colors from \citet{popescu_2018} and transformations from \citet{stassun_2018}, then added an estimate of the star's baseline flux using its $T$\,mag and the zero-point conversion from \citet{takacs_2025}. Finally, we again queried the postage stamp header to determine which pixels were used in the ``background'' aperture and which were used in the ``target'' aperture. We summed up the model flux in each of these apertures to produce a model SAP light curve.

Figures \ref{fig:nearest_asteroid_frame} and \ref{fig:lightcurve_contamination} show the result of this procedure and its agreement with real TESS SAP data. We note that the \jorbit model is a pure post-diction, not a fit to the data (aside from a baseline correction): the slight underestimates of the amplitude of the disturbance is likely a result of our crude color conversion.

\subsection{Simulating the 2029 flyby of (99942) Apophis} \label{sub:apophis}

As a final demonstration of \jorbit's ability to synthesize precomputed ephemeris data and a high-precision integrator, we simulated the April 2029 close approach of the asteroid (99942) Apophis. To accurately predict the outcome of Apophis' deflection, one must account for relativistic effects, non-gravitational acceleration, and gravitational harmonics of the Earth and Sun. This challenging situation has received considerable attention from the dynamics community, and was used as a test case in \assist's release paper and online documentation.

In addition to the usual JPL Horizons interface that is capable of generating and ephemeris for any given minor planet, the JPL Solar System Dynamics group (SSD) has produced several Apophis-specific ephemerides primarily in support of the OSIRIS-APEX mission. These models use smaller time steps/time spans and explicitly include recent stellar occultations reported to the Minor Planet Center. The most recent of these models, named ``orbit solution 220'', was released in Fall 2024 and is available in the SPK transfer format on the SSD's public FTP file browser\footnote{See memo \href{https://ssd.jpl.nasa.gov/ftp/eph/small_bodies/apophis/IOM392R-24-10-16-085745.pdf}{IOM 392R-24-10-16-085745} for additional information.}

For our simulation, we use the non-gravitational acceleration model of \citet{marsden_1973}, which is parameterized by a set of three $A$ coefficients for asteroids. We adopt the values of $A_1 = 5e-13, A_2 = -2.901e-14, A_3 = 0.0$, matching those from JPL's solution 220. For our gravitational harmonics model, we use the $J_2$ moment of the Sun (2.196e-7), and the $J_2, J_3, $ and $J_4$ moments of Earth (1.08e-03, -2.53e-06, and -1.62e-06, respectively). Our gravitational harmonics are assumed to be additional Newtonian contributions to the total acceleration, and we do not include relativistic corrections. Further, we do not self-consistently model the orientation of the spin axis of the Earth and Sun \citep[unlike the method in][]{lu_spin_2023}, and instead leave both fixed to their Horizons-predicted values for April 13th, 2029.

\begin{figure*}
    \centering
    \includegraphics[width=0.8\textwidth]{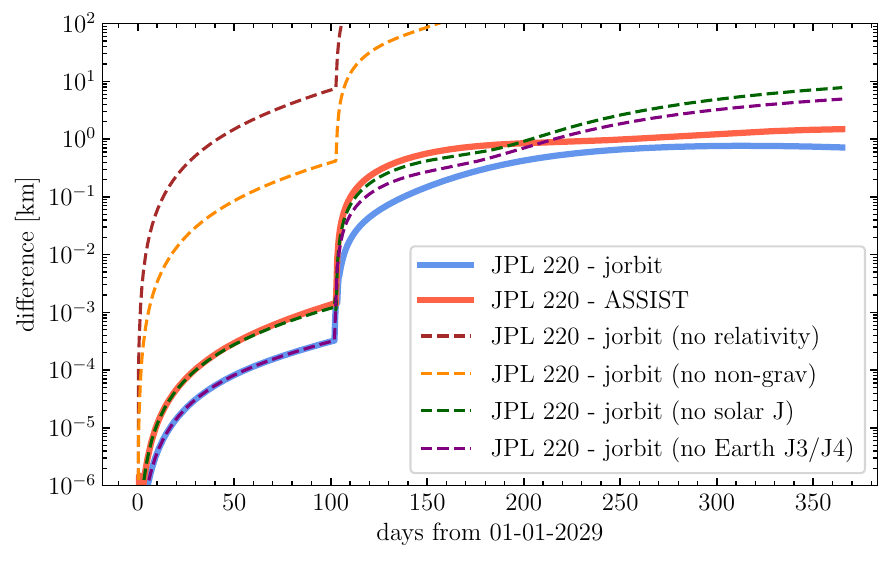}
    \caption{The difference in positions predicted by JPL's solution 220, \assist, and \jorbit. All simulations began from the same initial position/velocity on 01-01-2029 and their differences are the result of different implementations of the acceleration model and/or underlying integrator. Dashed lines show the results when leaving out individual components of the acceleration model. \pycodelink{https://github.com/ben-cassese/jorbit/blob/907e5b8ccee42479bf4fa67435b5733318d0bd9d/paper/apophis.ipynb}}
    \label{fig:apophis}
\end{figure*}

Our results are shown in Fig. \ref{fig:apophis}. Both \assist and \jorbit capably reproduce the JPL solution, with model packages agreeing within $\sim$1\,cm pre-encounter and within $\sim$1\,km by the end of the year. For comparison, the difference between JPL solution 220 and its predecessor, solution 218 from November of 2023, diverge by several hundred km over the same time range.

In addition to the default simulation that was based on the acceleration model described above, we also included several simulations that leave out individual acceleration components, which allows us to assess the relative importance of each. For example, we see that by neglecting relativistic corrections, the particle has already diverged more than a km even before the close encounter, leading to greatly enhanced post-flyby errors. The next most important contribution is the non-gravitational accelerations, since neglecting these similarly places the particle in the incorrect position pre-flyby. After these, the remaining contributions only affect the final difference in position at the $\sim$\,km level. Neglecting the solar $J_2$ moment produces a $\sim$cm shift pre-encounter, which translates to larger errors by the end of the year, while neglecting the Earth's $J_3$ and $J_4$ moments produces no distinguishable difference until the moment of closest approach.

\section{Conclusion} \label{sec:conclusion}

This publication marks our initial release of \jorbit, a framework that we hope will continue to develop and grow to meet data-driven challenges in solar system dynamics work. The core building blocks are now in place to expand in numerous possible directions. To name a few, \jorbit's foundations in the \jax ecosystem tempts an investigation into machine learning-accelerated inference; its flexible handling of acceleration functions suggests that it could be used to search for subtle effects like Yarkovsky acceleration; its general-purpose and differentiable integrator could be applied to other stellar systems to simulate exoplanet dynamics, potentially for searching the highly multi-modal space of transit timing variations \citep{flowmc, yahalomi_2024}.

In its current form, however, \jorbit already fills several previously unmet needs: it provides an open source autodifferentiable implementation of a high-precision integrator, and the ability to identify (and pre/postdict) minor planets near a given coordinate at a given time via an easy-to-use, locally-run software package. We look forward to seeing how the minor planet and broader dynamics community use these tools in the future.

\vspace{1cm}
We thank the IvyPlus Exchange Scholar Program for facilitating a long-term visit by B.C. in the Yale Department of Astronomy during part of this project. We thank Hanno Rein and an additional anonymous reviewer for their comments that greatly improved this manuscript and the code described within. This work was supported by NASA TESS GI grant \#80NSSC24K0359 and Heising-Simons Foundation Grants \#2021-2802 and \#2023-4478.
Data from the MPC's database is made freely available to the public. Funding for this data and the MPC's operations comes from a NASA PDCO grant (80NSSC22M0024), administered via a University of Maryland - SAO subaward (106075-Z6415201). The MPC's computing equipment is funded in part by the above award, and in part by funding from the Tamkin Foundation.
This research has made use of the Astrophysics Data System, funded by NASA under Cooperative Agreement 80NSSC21M00561.
We thank the Yale Center for Research Computing for guidance and assistance in using the Grace cluster.
This work used Anvil at Purdue University through allocation PHY250025 from the Advanced Cyberinfrastructure Coordination Ecosystem: Services \& Support (ACCESS) program, which is supported by U.S. National Science Foundation grants \#2138259, \#2138286, \#2138307, \#2137603, and \#2138296.

\software{
    \assist \citep{holman_assist_2023}; \texttt{astropy} \citep{astropy:2013, astropy:2018, astropy:2022};  \texttt{astroquery} \citep{astroquery}; \jax \citep{jax}; \texttt{jplephem} \citep{jplephem}; \texttt{lightkurve} \citep{lightkurve}; \texttt{matplotlib} \citep{hunter2007matplotlib}; \texttt{numpy} \citep{oliphant2006guide, walt2011numpy, harris2020array}, \rebound \citep{rein_rebound_2012}; \texttt{REBOUNDx} \citep{reboundx}; \texttt{scipy} \citep{scipy}.
}

\appendix
\section{Verification of Autodifferentiation} \label{app:autodiff}

Here we take the opportunity to verify that \jorbit's gradient calculations as computed via automatic differentiation agree with those computed via a simple finite differencing routine. For this test, we used the same simulated data and best-fit solution from Sec. \ref{sub:wiki_fits}. We used \jorbit to  compute the gradient of the likelihood at the best-fit solution using its normal autodiff routines, then compared the results to what is achieved using finite differencing with progressively smaller step sizes.

As expected, the two methods converge with decreasing step size. At the smallest step sizes tested, the numerical derivatives began to break down due to floating point errors, which explains the small upturns in the subplots for true anomaly, inclination, longitude of ascending node, and argument of periapsis. Overall, however, this test demonstrates that \jorbit is capable of computing exact derivatives for any parameter in a dynamical model. We note that \jorbit's gradients came ``for free'', meaning that we did not hard-code any derivative rules. Further, we note that autodiff is not a feature of \jorbit, but of \jax itself.

\begin{figure*}
    \centering
    \includegraphics[width=\textwidth]{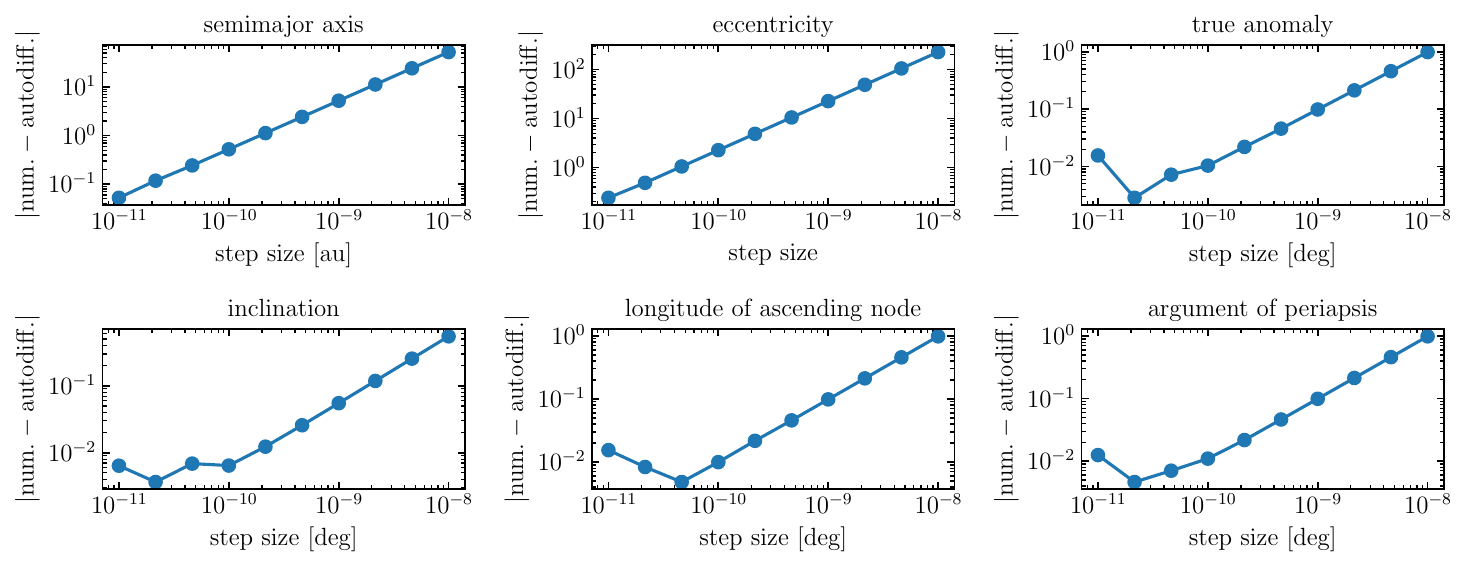}
    \caption{Comparison between \jorbit's derivative calculations computed via automatic differentiation (autodiff. on axes labels) and those computed by a simple finite differencing method (num. on axes labels). The upturn at small step sizes is due to numerical errors in the finite differencing scheme at small step sizes. \pycodelink{https://github.com/ben-cassese/jorbit/blob/352c8bf839ea5693f12f473c3f6c0574b4b0e16b/paper/autodiff_compare.ipynb}}
    \label{fig:autodiff_works}
\end{figure*}

\bibliography{references}{}

\begin{thebibliography}{}
\expandafter\ifx\csname natexlab\endcsname\relax\def\natexlab#1{#1}\fi
\providecommand{\url}[1]{\href{#1}{#1}}
\providecommand{\dodoi}[1]{doi:~\href{http://doi.org/#1}{\nolinkurl{#1}}}
\providecommand{\doeprint}[1]{\href{http://ascl.net/#1}{\nolinkurl{http://ascl.net/#1}}}
\providecommand{\doarXiv}[1]{\href{https://arxiv.org/abs/#1}{\nolinkurl{https://arxiv.org/abs/#1}}}

\bibitem[{{Acton}(1996)}]{acton_1996}
{Acton}, C.~H. 1996, \planss, 44, 65, \dodoi{10.1016/0032-0633(95)00107-7}

\bibitem[{{Annex} {et~al.}(2020){Annex}, {Pearson}, {Seignovert}, {Carcich},
  {Eichhorn}, {Mapel}, {von Forstner}, {McAuliffe}, {del Rio}, {Berry}, {Aye},
  {Stefko}, {de Val-Borro}, {Kulumani}, \& {Murakami}}]{annex_2020}
{Annex}, A., {Pearson}, B., {Seignovert}, B., {et~al.} 2020, The Journal of
  Open Source Software, 5, 2050, \dodoi{10.21105/joss.02050}

\bibitem[{{Arora} \& {Russell}(2010)}]{arora_2010}
{Arora}, N., \& {Russell}, R.~P. 2010, Celestial Mechanics and Dynamical
  Astronomy, 108, 107, \dodoi{10.1007/s10569-010-9296-0}

\bibitem[{{Astropy Collaboration} {et~al.}(2013){Astropy Collaboration},
  {Robitaille}, {Tollerud}, {Greenfield}, {Droettboom}, {Bray}, {Aldcroft},
  {Davis}, {Ginsburg}, {Price-Whelan}, {Kerzendorf}, {Conley}, {Crighton},
  {Barbary}, {Muna}, {Ferguson}, {Grollier}, {Parikh}, {Nair}, {Unther},
  {Deil}, {Woillez}, {Conseil}, {Kramer}, {Turner}, {Singer}, {Fox}, {Weaver},
  {Zabalza}, {Edwards}, {Azalee Bostroem}, {Burke}, {Casey}, {Crawford},
  {Dencheva}, {Ely}, {Jenness}, {Labrie}, {Lim}, {Pierfederici}, {Pontzen},
  {Ptak}, {Refsdal}, {Servillat}, \& {Streicher}}]{astropy:2013}
{Astropy Collaboration}, {Robitaille}, T.~P., {Tollerud}, E.~J., {et~al.} 2013,
  \aap, 558, A33, \dodoi{10.1051/0004-6361/201322068}

\bibitem[{{Astropy Collaboration} {et~al.}(2018){Astropy Collaboration},
  {Price-Whelan}, {Sip{\H{o}}cz}, {G{\"u}nther}, {Lim}, {Crawford}, {Conseil},
  {Shupe}, {Craig}, {Dencheva}, {Ginsburg}, {Vand erPlas}, {Bradley},
  {P{\'e}rez-Su{\'a}rez}, {de Val-Borro}, {Aldcroft}, {Cruz}, {Robitaille},
  {Tollerud}, {Ardelean}, {Babej}, {Bach}, {Bachetti}, {Bakanov}, {Bamford},
  {Barentsen}, {Barmby}, {Baumbach}, {Berry}, {Biscani}, {Boquien}, {Bostroem},
  {Bouma}, {Brammer}, {Bray}, {Breytenbach}, {Buddelmeijer}, {Burke},
  {Calderone}, {Cano Rodr{\'\i}guez}, {Cara}, {Cardoso}, {Cheedella}, {Copin},
  {Corrales}, {Crichton}, {D'Avella}, {Deil}, {Depagne}, {Dietrich}, {Donath},
  {Droettboom}, {Earl}, {Erben}, {Fabbro}, {Ferreira}, {Finethy}, {Fox},
  {Garrison}, {Gibbons}, {Goldstein}, {Gommers}, {Greco}, {Greenfield},
  {Groener}, {Grollier}, {Hagen}, {Hirst}, {Homeier}, {Horton}, {Hosseinzadeh},
  {Hu}, {Hunkeler}, {Ivezi{\'c}}, {Jain}, {Jenness}, {Kanarek}, {Kendrew},
  {Kern}, {Kerzendorf}, {Khvalko}, {King}, {Kirkby}, {Kulkarni}, {Kumar},
  {Lee}, {Lenz}, {Littlefair}, {Ma}, {Macleod}, {Mastropietro}, {McCully},
  {Montagnac}, {Morris}, {Mueller}, {Mumford}, {Muna}, {Murphy}, {Nelson},
  {Nguyen}, {Ninan}, {N{\"o}the}, {Ogaz}, {Oh}, {Parejko}, {Parley}, {Pascual},
  {Patil}, {Patil}, {Plunkett}, {Prochaska}, {Rastogi}, {Reddy Janga},
  {Sabater}, {Sakurikar}, {Seifert}, {Sherbert}, {Sherwood-Taylor}, {Shih},
  {Sick}, {Silbiger}, {Singanamalla}, {Singer}, {Sladen}, {Sooley},
  {Sornarajah}, {Streicher}, {Teuben}, {Thomas}, {Tremblay}, {Turner},
  {Terr{\'o}n}, {van Kerkwijk}, {de la Vega}, {Watkins}, {Weaver}, {Whitmore},
  {Woillez}, {Zabalza}, \& {Astropy Contributors}}]{astropy:2018}
{Astropy Collaboration}, {Price-Whelan}, A.~M., {Sip{\H{o}}cz}, B.~M., {et~al.}
  2018, \aj, 156, 123, \dodoi{10.3847/1538-3881/aabc4f}

\bibitem[{{Astropy Collaboration} {et~al.}(2022){Astropy Collaboration},
  {Price-Whelan}, {Lim}, {Earl}, {Starkman}, {Bradley}, {Shupe}, {Patil},
  {Corrales}, {Brasseur}, {N{"o}the}, {Donath}, {Tollerud}, {Morris},
  {Ginsburg}, {Vaher}, {Weaver}, {Tocknell}, {Jamieson}, {van Kerkwijk},
  {Robitaille}, {Merry}, {Bachetti}, {G{"u}nther}, {Aldcroft},
  {Alvarado-Montes}, {Archibald}, {B{'o}di}, {Bapat}, {Barentsen}, {Baz{'a}n},
  {Biswas}, {Boquien}, {Burke}, {Cara}, {Cara}, {Conroy}, {Conseil}, {Craig},
  {Cross}, {Cruz}, {D'Eugenio}, {Dencheva}, {Devillepoix}, {Dietrich},
  {Eigenbrot}, {Erben}, {Ferreira}, {Foreman-Mackey}, {Fox}, {Freij}, {Garg},
  {Geda}, {Glattly}, {Gondhalekar}, {Gordon}, {Grant}, {Greenfield}, {Groener},
  {Guest}, {Gurovich}, {Handberg}, {Hart}, {Hatfield-Dodds}, {Homeier},
  {Hosseinzadeh}, {Jenness}, {Jones}, {Joseph}, {Kalmbach}, {Karamehmetoglu},
  {Ka{l}uszy{'n}ski}, {Kelley}, {Kern}, {Kerzendorf}, {Koch}, {Kulumani},
  {Lee}, {Ly}, {Ma}, {MacBride}, {Maljaars}, {Muna}, {Murphy}, {Norman},
  {O'Steen}, {Oman}, {Pacifici}, {Pascual}, {Pascual-Granado}, {Patil},
  {Perren}, {Pickering}, {Rastogi}, {Roulston}, {Ryan}, {Rykoff}, {Sabater},
  {Sakurikar}, {Salgado}, {Sanghi}, {Saunders}, {Savchenko}, {Schwardt},
  {Seifert-Eckert}, {Shih}, {Jain}, {Shukla}, {Sick}, {Simpson},
  {Singanamalla}, {Singer}, {Singhal}, {Sinha}, {Sip{H{o}}cz}, {Spitler},
  {Stansby}, {Streicher}, {{{S}}umak}, {Swinbank}, {Taranu}, {Tewary},
  {Tremblay}, {Val-Borro}, {Van Kooten}, {Vasovi{'c}}, {Verma}, {de Miranda
  Cardoso}, {Williams}, {Wilson}, {Winkel}, {Wood-Vasey}, {Xue}, {Yoachim},
  {Zhang}, {Zonca}, \& {Astropy Project Contributors}}]{astropy:2022}
{Astropy Collaboration}, {Price-Whelan}, A.~M., {Lim}, P.~L., {et~al.} 2022,
  \apj, 935, 167, \dodoi{10.3847/1538-4357/ac7c74}

\bibitem[{{Bernstein} \& {Khushalani}(2000)}]{bernstein_2000}
{Bernstein}, G., \& {Khushalani}, B. 2000, \aj, 120, 3323,
  \dodoi{10.1086/316868}

\bibitem[{Boerner {et~al.}(2023)Boerner, Deems, Furlani, Knuth, \&
  Towns}]{nsf_access}
Boerner, T.~J., Deems, S., Furlani, T.~R., Knuth, S.~L., \& Towns, J. 2023, in
  Practice and Experience in Advanced Research Computing 2023: Computing for
  the Common Good, PEARC '23 (New York, NY, USA: Association for Computing
  Machinery), 173–176, \dodoi{10.1145/3569951.3597559}

\bibitem[{Bradbury {et~al.}(2018)Bradbury, Frostig, Hawkins, Johnson, Leary,
  Maclaurin, Necula, Paszke, Vander{P}las, Wanderman-{M}ilne, \& Zhang}]{jax}
Bradbury, J., Frostig, R., Hawkins, P., {et~al.} 2018, {JAX}: composable
  transformations of {P}ython+{N}um{P}y programs, 0.3.13.
\newblock \url{http://github.com/jax-ml/jax}

\bibitem[{{Cassese} {et~al.}(2025){Cassese}, {Lu}, \& Malena}]{zenodo_release}
{Cassese}, B., {Lu}, T., \& Malena, R. 2025, {jorbit: v1.0.0}, v1.0.0,  Zenodo,
  \dodoi{10.5281/zenodo.15843748}

\bibitem[{Dekker(1971)}]{dekker_1971}
Dekker, T.~J. 1971, Numerische Mathematik, 18, 224

\bibitem[{{Duncan} \& {Lissauer}(1998)}]{duncan_1998}
{Duncan}, M.~J., \& {Lissauer}, J.~J. 1998, \icarus, 134, 303,
  \dodoi{10.1006/icar.1998.5962}

\bibitem[{{Everhart}(1985)}]{everhart_1985}
{Everhart}, E. 1985, in Astrophysics and Space Science Library, Vol. 115, IAU
  Colloq. 83: Dynamics of Comets: Their Origin and Evolution, ed. A.~{Carusi}
  \& G.~B. {Valsecchi}, 185, \dodoi{10.1007/978-94-009-5400-7_17}

\bibitem[{{Fienga} {et~al.}(2016){Fienga}, {Laskar}, {Manche}, \&
  {Gastineau}}]{fienga_2016}
{Fienga}, A., {Laskar}, J., {Manche}, H., \& {Gastineau}, M. 2016, \aap, 587,
  L8, \dodoi{10.1051/0004-6361/201628227}

\bibitem[{{Fienga} {et~al.}(2008){Fienga}, {Manche}, {Laskar}, \&
  {Gastineau}}]{inpop06}
{Fienga}, A., {Manche}, H., {Laskar}, J., \& {Gastineau}, M. 2008, \aap, 477,
  315, \dodoi{10.1051/0004-6361:20066607}

\bibitem[{{Gaia Collaboration} {et~al.}(2018{\natexlab{a}}){Gaia
  Collaboration}, {Spoto}, {Tanga}, {Mignard}, {Berthier}, {Carry}, {Cellino},
  {Dell'Oro}, {Hestroffer}, {Muinonen}, {Pauwels}, {Petit}, {David}, {De
  Angeli}, {Delbo}, {Fr{\'e}zouls}, {Galluccio}, {Granvik}, {Guiraud},
  {Hern{\'a}ndez}, {Ord{\'e}novic}, {Portell}, {Poujoulet}, {Thuillot},
  {Walmsley}, {Brown}, {Vallenari}, {Prusti}, {de Bruijne}, {Babusiaux},
  {Bailer-Jones}, {Biermann}, {Evans}, {Eyer}, {Jansen}, {Jordi}, {Klioner},
  {Lammers}, {Lindegren}, {Luri}, {Panem}, {Pourbaix}, {Randich}, {Sartoretti},
  {Siddiqui}, {Soubiran}, {van Leeuwen}, {Walton}, {Arenou}, {Bastian},
  {Cropper}, {Drimmel}, {Katz}, {Lattanzi}, {Bakker}, {Cacciari},
  {Casta{\~n}eda}, {Chaoul}, {Cheek}, {Fabricius}, {Guerra}, {Holl}, {Masana},
  {Messineo}, {Mowlavi}, {Nienartowicz}, {Panuzzo}, {Riello}, {Seabroke},
  {Th{\'e}venin}, {Gracia-Abril}, {Comoretto}, {Garcia-Reinaldos}, {Teyssier},
  {Altmann}, {Andrae}, {Audard}, {Bellas-Velidis}, {Benson}, {Blomme},
  {Burgess}, {Busso}, {Clementini}, {Clotet}, {Creevey}, {Davidson}, {De
  Ridder}, {Delchambre}, {Ducourant}, {Fern{\'a}ndez-Hern{\'a}ndez},
  {Fouesneau}, {Fr{\'e}mat}, {Garc{\'\i}a-Torres},
  {Gonz{\'a}lez-N{\'u}{\~n}ez}, {Gonz{\'a}lez-Vidal}, {Gosset}, {Guy},
  {Halbwachs}, {Hambly}, {Harrison}, {Hodgkin}, {Hutton}, {Jasniewicz},
  {Jean-Antoine-Piccolo}, {Jordan}, {Korn}, {Krone-Martins}, {Lanzafame},
  {Lebzelter}, {L{\"o}}, {Manteiga}, {Marrese}, {Mart{\'\i}n-Fleitas},
  {Moitinho}, {Mora}, {Osinde}, {Pancino}, {Recio-Blanco}, {Richards},
  {Rimoldini}, {Robin}, {Sarro}, {Siopis}, {Smith}, {Sozzetti}, {S{\"u}veges},
  {Torra}, {van Reeven}, {Abbas}, {Abreu Aramburu}, {Accart}, {Aerts},
  {Altavilla}, {{\'A}lvarez}, {Alvarez}, {Alves}, {Anderson}, {Andrei},
  {Anglada Varela}, {Antiche}, {Antoja}, {Arcay}, {Astraatmadja}, {Bach},
  {Baker}, {Balaguer-N{\'u}{\~n}ez}, {Balm}, {Barache}, {Barata}, {Barbato},
  {Barblan}, {Barklem}, {Barrado}, {Barros}, {Barstow}, {Bartholom{\'e}
  Mu{\~n}oz}, {Bassilana}, {Becciani}, {Bellazzini}, {Berihuete}, {Bertone},
  {Bianchi}, {Bienaym{\'e}}, {Blanco-Cuaresma}, {Boch}, {Boeche}, {Bombrun},
  {Borrachero}, {Bossini}, {Bouquillon}, {Bourda}, {Bragaglia}, {Bramante},
  {Breddels}, {Bressan}, {Brouillet}, {Br{\"u}semeister}, {Brugaletta},
  {Bucciarelli}, {Burlacu}, {Busonero}, {Butkevich}, {Buzzi}, {Caffau},
  {Cancelliere}, {Cannizzaro}, {Cantat-Gaudin}, {Carballo}, {Carlucci},
  {Carrasco}, {Casamiquela}, {Castellani}, {Castro-Ginard}, {Charlot},
  {Chemin}, {Chiavassa}, {Cocozza}, {Costigan}, {Cowell}, \&
  {Crifo}}]{gaia_dr2_solar_system_2018}
{Gaia Collaboration}, {Spoto}, F., {Tanga}, P., {et~al.} 2018{\natexlab{a}},
  \aap, 616, A13, \dodoi{10.1051/0004-6361/201832900}

\bibitem[{{Gaia Collaboration} {et~al.}(2018{\natexlab{b}}){Gaia
  Collaboration}, {Spoto}, {Tanga}, {Mignard}, {Berthier}, {Carry}, {Cellino},
  {Dell'Oro}, {Hestroffer}, {Muinonen}, {Pauwels}, {Petit}, {David}, {De
  Angeli}, {Delbo}, {Fr{\'e}zouls}, {Galluccio}, {Granvik}, {Guiraud},
  {Hern{\'a}ndez}, {Ord{\'e}novic}, {Portell}, {Poujoulet}, {Thuillot},
  {Walmsley}, {Brown}, {Vallenari}, {Prusti}, {de Bruijne}, {Babusiaux},
  {Bailer-Jones}, {Biermann}, {Evans}, {Eyer}, {Jansen}, {Jordi}, {Klioner},
  {Lammers}, {Lindegren}, {Luri}, {Panem}, {Pourbaix}, {Randich}, {Sartoretti},
  {Siddiqui}, {Soubiran}, {van Leeuwen}, {Walton}, {Arenou}, {Bastian},
  {Cropper}, {Drimmel}, {Katz}, {Lattanzi}, {Bakker}, {Cacciari},
  {Casta{\~n}eda}, {Chaoul}, {Cheek}, {Fabricius}, {Guerra}, {Holl}, {Masana},
  {Messineo}, {Mowlavi}, {Nienartowicz}, {Panuzzo}, {Riello}, {Seabroke},
  {Th{\'e}venin}, {Gracia-Abril}, {Comoretto}, {Garcia-Reinaldos}, {Teyssier},
  {Altmann}, {Andrae}, {Audard}, {Bellas-Velidis}, {Benson}, {Blomme},
  {Burgess}, {Busso}, {Clementini}, {Clotet}, {Creevey}, {Davidson}, {De
  Ridder}, {Delchambre}, {Ducourant}, {Fern{\'a}ndez-Hern{\'a}ndez},
  {Fouesneau}, {Fr{\'e}mat}, {Garc{\'\i}a-Torres},
  {Gonz{\'a}lez-N{\'u}{\~n}ez}, {Gonz{\'a}lez-Vidal}, {Gosset}, {Guy},
  {Halbwachs}, {Hambly}, {Harrison}, {Hodgkin}, {Hutton}, {Jasniewicz},
  {Jean-Antoine-Piccolo}, {Jordan}, {Korn}, {Krone-Martins}, {Lanzafame},
  {Lebzelter}, {L{\"o}}, {Manteiga}, {Marrese}, {Mart{\'\i}n-Fleitas},
  {Moitinho}, {Mora}, {Osinde}, {Pancino}, {Recio-Blanco}, {Richards},
  {Rimoldini}, {Robin}, {Sarro}, {Siopis}, {Smith}, {Sozzetti}, {S{\"u}veges},
  {Torra}, {van Reeven}, {Abbas}, {Abreu Aramburu}, {Accart}, {Aerts},
  {Altavilla}, {{\'A}lvarez}, {Alvarez}, {Alves}, {Anderson}, {Andrei},
  {Anglada Varela}, {Antiche}, {Antoja}, {Arcay}, {Astraatmadja}, {Bach},
  {Baker}, {Balaguer-N{\'u}{\~n}ez}, {Balm}, {Barache}, {Barata}, {Barbato},
  {Barblan}, {Barklem}, {Barrado}, {Barros}, {Barstow}, {Bartholom{\'e}
  Mu{\~n}oz}, {Bassilana}, {Becciani}, {Bellazzini}, {Berihuete}, {Bertone},
  {Bianchi}, {Bienaym{\'e}}, {Blanco-Cuaresma}, {Boch}, {Boeche}, {Bombrun},
  {Borrachero}, {Bossini}, {Bouquillon}, {Bourda}, {Bragaglia}, {Bramante},
  {Breddels}, {Bressan}, {Brouillet}, {Br{\"u}semeister}, {Brugaletta},
  {Bucciarelli}, {Burlacu}, {Busonero}, {Butkevich}, {Buzzi}, {Caffau},
  {Cancelliere}, {Cannizzaro}, {Cantat-Gaudin}, {Carballo}, {Carlucci},
  {Carrasco}, {Casamiquela}, {Castellani}, {Castro-Ginard}, {Charlot},
  {Chemin}, {Chiavassa}, {Cocozza}, {Costigan}, {Cowell}, \&
  {Crifo}}]{spoto_2018}
---. 2018{\natexlab{b}}, \aap, 616, A13, \dodoi{10.1051/0004-6361/201832900}

\bibitem[{{Gauss}(1809)}]{gauss_1809}
{Gauss}, K.~F. 1809, Theoria motvs corporvm coelestivm in sectionibvs conicis
  solem ambientivm

\bibitem[{{Ginsburg} {et~al.}(2019){Ginsburg}, {Sip{\H{o}}cz}, {Brasseur},
  {Cowperthwaite}, {Craig}, {Deil}, {Guillochon}, {Guzman}, {Liedtke}, {Lian
  Lim}, {Lockhart}, {Mommert}, {Morris}, {Norman}, {Parikh}, {Persson},
  {Robitaille}, {Segovia}, {Singer}, {Tollerud}, {de Val-Borro}, {Valtchanov},
  {Woillez}, {Astroquery Collaboration}, \& {a subset of astropy
  Collaboration}}]{astroquery}
{Ginsburg}, A., {Sip{\H{o}}cz}, B.~M., {Brasseur}, C.~E., {et~al.} 2019, \aj,
  157, 98, \dodoi{10.3847/1538-3881/aafc33}

\bibitem[{Harris {et~al.}(2020)Harris, Millman, van~der Walt, Gommers,
  Virtanen, Cournapeau, Wieser, Taylor, Berg, Smith,
  {et~al.}}]{harris2020array}
Harris, C.~R., Millman, K.~J., van~der Walt, S.~J., {et~al.} 2020, Nature, 585,
  357

\bibitem[{{Herget}(1965)}]{herget_1965}
{Herget}, P. 1965, \aj, 70, 1, \dodoi{10.1086/109671}

\bibitem[{{Holman} {et~al.}(2023){Holman}, {Akmal}, {Farnocchia}, {Rein},
  {Payne}, {Weryk}, {Tamayo}, \& {Hernandez}}]{holman_assist_2023}
{Holman}, M.~J., {Akmal}, A., {Farnocchia}, D., {et~al.} 2023, \psj, 4, 69,
  \dodoi{10.3847/PSJ/acc9a9}

\bibitem[{{Howell} {et~al.}(2014){Howell}, {Sobeck}, {Haas}, {Still},
  {Barclay}, {Mullally}, {Troeltzsch}, {Aigrain}, {Bryson}, {Caldwell},
  {Chaplin}, {Cochran}, {Huber}, {Marcy}, {Miglio}, {Najita}, {Smith},
  {Twicken}, \& {Fortney}}]{k2}
{Howell}, S.~B., {Sobeck}, C., {Haas}, M., {et~al.} 2014, \pasp, 126, 398,
  \dodoi{10.1086/676406}

\bibitem[{Hunter(2007)}]{hunter2007matplotlib}
Hunter, J.~D. 2007, Computing in science \& engineering, 9, 90

\bibitem[{{Ivezi{\'c}} {et~al.}(2019){Ivezi{\'c}}, {Kahn}, {Tyson}, {Abel},
  {Acosta}, {Allsman}, {Alonso}, {AlSayyad}, {Anderson}, {Andrew}, {Angel},
  {Angeli}, {Ansari}, {Antilogus}, {Araujo}, {Armstrong}, {Arndt}, {Astier},
  {Aubourg}, {Auza}, {Axelrod}, {Bard}, {Barr}, {Barrau}, {Bartlett}, {Bauer},
  {Bauman}, {Baumont}, {Bechtol}, {Bechtol}, {Becker}, {Becla}, {Beldica},
  {Bellavia}, {Bianco}, {Biswas}, {Blanc}, {Blazek}, {Blandford}, {Bloom},
  {Bogart}, {Bond}, {Booth}, {Borgland}, {Borne}, {Bosch}, {Boutigny},
  {Brackett}, {Bradshaw}, {Brandt}, {Brown}, {Bullock}, {Burchat}, {Burke},
  {Cagnoli}, {Calabrese}, {Callahan}, {Callen}, {Carlin}, {Carlson},
  {Chandrasekharan}, {Charles-Emerson}, {Chesley}, {Cheu}, {Chiang}, {Chiang},
  {Chirino}, {Chow}, {Ciardi}, {Claver}, {Cohen-Tanugi}, {Cockrum}, {Coles},
  {Connolly}, {Cook}, {Cooray}, {Covey}, {Cribbs}, {Cui}, {Cutri}, {Daly},
  {Daniel}, {Daruich}, {Daubard}, {Daues}, {Dawson}, {Delgado}, {Dellapenna},
  {de Peyster}, {de Val-Borro}, {Digel}, {Doherty}, {Dubois},
  {Dubois-Felsmann}, {Durech}, {Economou}, {Eifler}, {Eracleous}, {Emmons},
  {Fausti Neto}, {Ferguson}, {Figueroa}, {Fisher-Levine}, {Focke}, {Foss},
  {Frank}, {Freemon}, {Gangler}, {Gawiser}, {Geary}, {Gee}, {Geha}, {Gessner},
  {Gibson}, {Gilmore}, {Glanzman}, {Glick}, {Goldina}, {Goldstein}, {Goodenow},
  {Graham}, {Gressler}, {Gris}, {Guy}, {Guyonnet}, {Haller}, {Harris},
  {Hascall}, {Haupt}, {Hernandez}, {Herrmann}, {Hileman}, {Hoblitt}, {Hodgson},
  {Hogan}, {Howard}, {Huang}, {Huffer}, {Ingraham}, {Innes}, {Jacoby}, {Jain},
  {Jammes}, {Jee}, {Jenness}, {Jernigan}, {Jevremovi{\'c}}, {Johns}, {Johnson},
  {Johnson}, {Jones}, {Juramy-Gilles}, {Juri{\'c}}, {Kalirai}, {Kallivayalil},
  {Kalmbach}, {Kantor}, {Karst}, {Kasliwal}, {Kelly}, {Kessler}, {Kinnison},
  {Kirkby}, {Knox}, {Kotov}, {Krabbendam}, {Krughoff}, {Kub{\'a}nek},
  {Kuczewski}, {Kulkarni}, {Ku}, {Kurita}, {Lage}, {Lambert}, {Lange},
  {Langton}, {Le Guillou}, {Levine}, {Liang}, {Lim}, {Lintott}, {Long},
  {Lopez}, {Lotz}, {Lupton}, {Lust}, {MacArthur}, {Mahabal}, {Mandelbaum},
  {Markiewicz}, {Marsh}, {Marshall}, {Marshall}, {May}, {McKercher}, {McQueen},
  {Meyers}, {Migliore}, {Miller}, \& {Mills}}]{lsst}
{Ivezi{\'c}}, {\v{Z}}., {Kahn}, S.~M., {Tyson}, J.~A., {et~al.} 2019, \apj,
  873, 111, \dodoi{10.3847/1538-4357/ab042c}

\bibitem[{{Jenkins} {et~al.}(2016){Jenkins}, {Twicken}, {McCauliff},
  {Campbell}, {Sanderfer}, {Lung}, {Mansouri-Samani}, {Girouard}, {Tenenbaum},
  {Klaus}, {Smith}, {Caldwell}, {Chacon}, {Henze}, {Heiges}, {Latham},
  {Morgan}, {Swade}, {Rinehart}, \& {Vanderspek}}]{tess_spoc}
{Jenkins}, J.~M., {Twicken}, J.~D., {McCauliff}, S., {et~al.} 2016, in Society
  of Photo-Optical Instrumentation Engineers (SPIE) Conference Series, Vol.
  9913, Software and Cyberinfrastructure for Astronomy IV, ed. G.~{Chiozzi} \&
  J.~C. {Guzman}, 99133E, \dodoi{10.1117/12.2233418}

\bibitem[{{Johansson} {et~al.}(2017){Johansson}, {Steinberg}, {Kirpichev},
  {Kuhlman}, {Meurer}, {{\v{C}}ert{\'\i}k}, {Van Horsen}, {Masson}, {Arias De
  Reyna}, {Hartmann}, {Pernici}, {Kagalenko}, {Peterson},
  {J{\k{e}}drzejewski-Szmek}, {Krastanov}, {Warner}, {Weckesser}, {Buchert},
  {Schl{\"o}mer}, {Creus-Costa}, {Ingold}, {Behan}, \& {Brys}}]{mpmath}
{Johansson}, F., {Steinberg}, V., {Kirpichev}, S.~B., {et~al.} 2017, {mpmath: a
  Python library for arbitrary-precision floating-point arithmetic}, 1.0.0,
  Zenodo, \dodoi{10.5281/zenodo.1476881}

\bibitem[{Kent(1982)}]{kent_1982}
Kent, J.~T. 1982, Journal of the Royal Statistical Society: Series B
  (Methodological), 44, 71

\bibitem[{{Lightkurve Collaboration} {et~al.}(2018){Lightkurve Collaboration},
  {Cardoso}, {Hedges}, {Gully-Santiago}, {Saunders}, {Cody}, {Barclay}, {Hall},
  {Sagear}, {Turtelboom}, {Zhang}, {Tzanidakis}, {Mighell}, {Coughlin}, {Bell},
  {Berta-Thompson}, {Williams}, {Dotson}, \& {Barentsen}}]{lightkurve}
{Lightkurve Collaboration}, {Cardoso}, J.~V.~d.~M., {Hedges}, C., {et~al.}
  2018, {Lightkurve: Kepler and TESS time series analysis in Python},
  Astrophysics Source Code Library.
\newblock \doeprint{1812.013}

\bibitem[{{Lu} {et~al.}(2023){Lu}, {Rein}, {Tamayo}, {Hadden}, {Mardling},
  {Millholland}, \& {Laughlin}}]{lu_spin_2023}
{Lu}, T., {Rein}, H., {Tamayo}, D., {et~al.} 2023, \apj, 948, 41,
  \dodoi{10.3847/1538-4357/acc06d}

\bibitem[{{Makadia} {et~al.}(2023){Makadia}, {Chesley}, {Farnocchia}, \&
  {Eggl}}]{grss}
{Makadia}, R., {Chesley}, S., {Farnocchia}, D., \& {Eggl}, S. 2023, in
  AAS/Division for Planetary Sciences Meeting Abstracts, Vol.~55, AAS/Division
  for Planetary Sciences Meeting Abstracts \#55, 405.02

\bibitem[{{Marsden}(1985)}]{marsden_1985}
{Marsden}, B.~G. 1985, \aj, 90, 1541, \dodoi{10.1086/113867}

\bibitem[{{Marsden} {et~al.}(1973){Marsden}, {Sekanina}, \&
  {Yeomans}}]{marsden_1973}
{Marsden}, B.~G., {Sekanina}, Z., \& {Yeomans}, D.~K. 1973, \aj, 78, 211,
  \dodoi{10.1086/111402}

\bibitem[{{Milani}(1999)}]{milani_1999}
{Milani}, A. 1999, \icarus, 137, 269, \dodoi{10.1006/icar.1999.6045}

\bibitem[{{Muinonen} \& {Bowell}(1993)}]{muinonen_1993}
{Muinonen}, K., \& {Bowell}, E. 1993, \icarus, 104, 255,
  \dodoi{10.1006/icar.1993.1100}

\bibitem[{{Newhall} {et~al.}(1983){Newhall}, {Standish}, \&
  {Williams}}]{newhall_1983}
{Newhall}, X.~X., {Standish}, E.~M., \& {Williams}, J.~G. 1983, \aap, 125, 150

\bibitem[{{Nobili} \& {Roxburgh}(1986)}]{nobili_1986}
{Nobili}, A., \& {Roxburgh}, I.~W. 1986, in IAU Symposium, Vol. 114, Relativity
  in Celestial Mechanics and Astrometry. High Precision Dynamical Theories and
  Observational Verifications, ed. J.~{Kovalevsky} \& V.~A. {Brumberg}, 105

\bibitem[{Oliphant(2006)}]{oliphant2006guide}
Oliphant, T.~E. 2006, A guide to NumPy, Vol.~1 (Trelgol Publishing USA)

\bibitem[{{Park} {et~al.}(2021){Park}, {Folkner}, {Williams}, \&
  {Boggs}}]{park_2021}
{Park}, R.~S., {Folkner}, W.~M., {Williams}, J.~G., \& {Boggs}, D.~H. 2021,
  \aj, 161, 105, \dodoi{10.3847/1538-3881/abd414}

\bibitem[{{Pham} {et~al.}(2024){Pham}, {Rein}, \&
  {Spiegel}}]{pham_timestep_2024}
{Pham}, D., {Rein}, H., \& {Spiegel}, D.~S. 2024, The Open Journal of
  Astrophysics, 7, 1, \dodoi{10.21105/astro.2401.02849}

\bibitem[{{Popescu} {et~al.}(2018){Popescu}, {Licandro}, {Carvano},
  {Stoicescu}, {de Le{\'o}n}, {Morate}, {Boac{\u{a}}}, \&
  {Cristescu}}]{popescu_2018}
{Popescu}, M., {Licandro}, J., {Carvano}, J.~M., {et~al.} 2018, \aap, 617, A12,
  \dodoi{10.1051/0004-6361/201833023}

\bibitem[{{Rein} \& {Liu}(2012)}]{rein_rebound_2012}
{Rein}, H., \& {Liu}, S.~F. 2012, \aap, 537, A128,
  \dodoi{10.1051/0004-6361/201118085}

\bibitem[{{Rein} \& {Spiegel}(2015)}]{rein_ias15_2015}
{Rein}, H., \& {Spiegel}, D.~S. 2015, \mnras, 446, 1424,
  \dodoi{10.1093/mnras/stu2164}

\bibitem[{{Rein} \& {Tamayo}(2016)}]{rein_2nd_order_2016}
{Rein}, H., \& {Tamayo}, D. 2016, \mnras, 459, 2275,
  \dodoi{10.1093/mnras/stw644}

\bibitem[{{Rhodes} {et~al.}(2019){Rhodes}, {van Kerkwijk}, {Davies},
  {Eichhorn}, \& {Rodr{\'\i}guez}}]{jplephem}
{Rhodes}, B., {van Kerkwijk}, M., {Davies}, J., {Eichhorn}, H., \&
  {Rodr{\'\i}guez}, J. 2019, {JPLephem: Jet Propulsion Lab ephemerides
  package}, Astrophysics Source Code Library, record ascl:1908.017

\bibitem[{{Ricker} {et~al.}(2015){Ricker}, {Winn}, {Vanderspek}, {Latham},
  {Bakos}, {Bean}, {Berta-Thompson}, {Brown}, {Buchhave}, {Butler}, {Butler},
  {Chaplin}, {Charbonneau}, {Christensen-Dalsgaard}, {Clampin}, {Deming},
  {Doty}, {De Lee}, {Dressing}, {Dunham}, {Endl}, {Fressin}, {Ge}, {Henning},
  {Holman}, {Howard}, {Ida}, {Jenkins}, {Jernigan}, {Johnson}, {Kaltenegger},
  {Kawai}, {Kjeldsen}, {Laughlin}, {Levine}, {Lin}, {Lissauer}, {MacQueen},
  {Marcy}, {McCullough}, {Morton}, {Narita}, {Paegert}, {Palle}, {Pepe},
  {Pepper}, {Quirrenbach}, {Rinehart}, {Sasselov}, {Sato}, {Seager},
  {Sozzetti}, {Stassun}, {Sullivan}, {Szentgyorgyi}, {Torres}, {Udry}, \&
  {Villasenor}}]{tess}
{Ricker}, G.~R., {Winn}, J.~N., {Vanderspek}, R., {et~al.} 2015, Journal of
  Astronomical Telescopes, Instruments, and Systems, 1, 014003,
  \dodoi{10.1117/1.JATIS.1.1.014003}

\bibitem[{Song {et~al.}(2022)Song, Smith, Kalyanam, Zhu, Adams, Colby,
  Finnegan, Gough, Hillery, Irvine, Maji, \& St.~John}]{anvil}
Song, X.~C., Smith, P., Kalyanam, R., {et~al.} 2022, in Practice and Experience
  in Advanced Research Computing 2022: Revolutionary: Computing, Connections,
  You, PEARC '22 (New York, NY, USA: Association for Computing Machinery),
  \dodoi{10.1145/3491418.3530766}

\bibitem[{{Stassun} {et~al.}(2018){Stassun}, {Oelkers}, {Pepper}, {Paegert},
  {De Lee}, {Torres}, {Latham}, {Charpinet}, {Dressing}, {Huber}, {Kane},
  {L{\'e}pine}, {Mann}, {Muirhead}, {Rojas-Ayala}, {Silvotti}, {Fleming},
  {Levine}, \& {Plavchan}}]{stassun_2018}
{Stassun}, K.~G., {Oelkers}, R.~J., {Pepper}, J., {et~al.} 2018, \aj, 156, 102,
  \dodoi{10.3847/1538-3881/aad050}

\bibitem[{{Stroud} \& {Secrest}(1966)}]{gaussian_quad_book}
{Stroud}, A., \& {Secrest}, D. 1966, {Gaussian quadrature formulas}
  ({Prentice-Hall}).
\newblock \url{https://lccn.loc.gov/66011596}

\bibitem[{{Tak{\'a}cs} {et~al.}(2025){Tak{\'a}cs}, {Kiss}, {Szak{\'a}ts}, \&
  {P{\'a}l}}]{takacs_2025}
{Tak{\'a}cs}, N., {Kiss}, C., {Szak{\'a}ts}, R., \& {P{\'a}l}, A. 2025, arXiv
  e-prints, arXiv:2503.13332, \dodoi{10.48550/arXiv.2503.13332}

\bibitem[{{Tamayo} {et~al.}(2020{\natexlab{a}}){Tamayo}, {Rein}, {Shi}, \&
  {Hernandez}}]{reboundx}
{Tamayo}, D., {Rein}, H., {Shi}, P., \& {Hernandez}, D.~M. 2020{\natexlab{a}},
  \mnras, 491, 2885, \dodoi{10.1093/mnras/stz2870}

\bibitem[{{Tamayo} {et~al.}(2020{\natexlab{b}}){Tamayo}, {Cranmer}, {Hadden},
  {Rein}, {Battaglia}, {Obertas}, {Armitage}, {Ho}, {Spergel}, {Gilbertson},
  {Hussain}, {Silburt}, {Jontof-Hutter}, \& {Menou}}]{tamayo_spock_2020}
{Tamayo}, D., {Cranmer}, M., {Hadden}, S., {et~al.} 2020{\natexlab{b}},
  Proceedings of the National Academy of Science, 117, 18194,
  \dodoi{10.1073/pnas.2001258117}

\bibitem[{{Tanga} {et~al.}(2023){Tanga}, {Pauwels}, {Mignard}, {Muinonen},
  {Cellino}, {David}, {Hestroffer}, {Spoto}, {Berthier}, {Guiraud}, {Roux},
  {Carry}, {Delbo}, {Dell'Oro}, {Fouron}, {Galluccio}, {Jonckheere}, {Klioner},
  {Lefustec}, {Liberato}, {Ord{\'e}novic}, {Oreshina-Slezak}, {Penttil{\"a}},
  {Pailler}, {Panem}, {Petit}, {Portell}, {Poujoulet}, {Thuillot}, {Van
  Hemelryck}, {Burlacu}, {Lasne}, \& {Managau}}]{tanga_gaia_dr3_2023}
{Tanga}, P., {Pauwels}, T., {Mignard}, F., {et~al.} 2023, \aap, 674, A12,
  \dodoi{10.1051/0004-6361/202243796}

\bibitem[{Virtanen {et~al.}(2020)Virtanen, Gommers, Oliphant, Haberland, Reddy,
  Cournapeau, Burovski, Peterson, Weckesser, Bright, {van der Walt}, Brett,
  Wilson, Millman, Mayorov, Nelson, Jones, Kern, Larson, Carey, Polat, Feng,
  Moore, {VanderPlas}, Laxalde, Perktold, Cimrman, Henriksen, Quintero, Harris,
  Archibald, Ribeiro, Pedregosa, {van Mulbregt}, \& {SciPy 1.0
  Contributors}}]{scipy}
Virtanen, P., Gommers, R., Oliphant, T.~E., {et~al.} 2020, Nature Methods, 17,
  261, \dodoi{10.1038/s41592-019-0686-2}

\bibitem[{Walt {et~al.}(2011)Walt, Colbert, \& Varoquaux}]{walt2011numpy}
Walt, S. v.~d., Colbert, S.~C., \& Varoquaux, G. 2011, Computing in Science \&
  Engineering, 13, 22

\bibitem[{{Wisdom} \& {Holman}(1991)}]{wisdom_holman_1991}
{Wisdom}, J., \& {Holman}, M. 1991, \aj, 102, 1528, \dodoi{10.1086/115978}

\bibitem[{{Wong} {et~al.}(2023){Wong}, {Gabri{\'e}}, \&
  {Foreman-Mackey}}]{flowmc}
{Wong}, K. W.~K., {Gabri{\'e}}, M., \& {Foreman-Mackey}, D. 2023, The Journal
  of Open Source Software, 8, 5021, \dodoi{10.21105/joss.05021}

\bibitem[{{Yahalomi} \& {Kipping}(2024)}]{yahalomi_2024}
{Yahalomi}, D.~A., \& {Kipping}, D. 2024, arXiv e-prints, arXiv:2411.10493,
  \dodoi{10.48550/arXiv.2411.10493}

\end{thebibliography}
\bibliographystyle{aasjournal}

\end{document}